\documentclass{article}

\usepackage{microtype}
\usepackage{graphicx}
\usepackage{subfigure}
\usepackage{booktabs} 
\usepackage{placeins}
\usepackage{diagbox}
\usepackage{multirow}
\usepackage{anyfontsize}

\usepackage{hyperref}

\usepackage[accepted]{icml2024}

\usepackage{amsmath}
\usepackage{amssymb}
\usepackage{mathtools}
\usepackage{amsthm}

\usepackage[capitalize,noabbrev]{cleveref}

\theoremstyle{plain}

\theoremstyle{definition}

\theoremstyle{remark}

\usepackage[textsize=tiny]{todonotes}

\icmltitlerunning{DiffDA: a Diffusion Model for Weather-scale Data Assimilation}

\usepackage{times}

\makeatletter
\newcommand{\HEADER}[1]{\ALC@it\underline{\textsc{#1}}\begin{ALC@g}}
\newcommand{\ENDHEADER}{\end{ALC@g}}
\makeatother

\renewcommand{\algorithmicrequire}{\textbf{Input:}}
\renewcommand{\algorithmicensure}{\textbf{Output:}}

\renewcommand{\headrulewidth}{0pt}

\usepackage{natbib}
\usepackage{hyperref}
\def\equationautorefname~#1\null{(#1)\null}

\begin{document}

\twocolumn[
\icmltitle{DiffDA: A Diffusion Model for Weather-scale Data Assimilation}



\icmlsetsymbol{equal}{*}

\begin{icmlauthorlist}
\icmlauthor{Langwen Huang}{ethz}
\icmlauthor{Lukas Gianinazzi}{ethz}
\icmlauthor{Yuejiang Yu}{ethz}
\icmlauthor{Peter D. Dueben}{ecmwf}
\icmlauthor{Torsten Hoefler}{ethz}
\end{icmlauthorlist}

\icmlaffiliation{ethz}{Department of Computer Science, ETH Zürich, Switzerland}
\icmlaffiliation{ecmwf}{European Centre for Medium-Range Weather Forecasts (ECMWF), Reading, United Kingdom}

\icmlcorrespondingauthor{Langwen Huang}{langwen.huang@inf.ethz.ch}
\icmlcorrespondingauthor{Torsten Hoefler}{torsten.hoefler@inf.ethz.ch}

\icmlkeywords{Data Assimilation, Denoising Diffusion Model}

\vskip 0.3in
]

\printAffiliationsAndNotice{\icmlEqualContribution} 


\begin{abstract}
The generation of initial conditions via accurate data assimilation is crucial for weather forecasting and climate modeling. We propose DiffDA as a denoising diffusion model capable of assimilating atmospheric variables using predicted states and sparse observations. Exploiting the similarity between a weather forecasting model and a denoising diffusion model dedicated to weather applications, we adapt the pretrained GraphCast neural network as the backbone of the diffusion model.  
Through experiments based on simulated observations from the ERA5 reanalysis dataset, our method can produce assimilated global atmospheric data consistent with observations at 0.25$^\circ$ ($\approx$30km) resolution globally. This marks the highest resolution achieved by ML data assimilation models. 
The experiments also show that the initial conditions assimilated from sparse observations (less than $0.96\%$ of gridded data) and 48-hour forecast can be used for forecast models 
with a loss of lead time of at most 24 hours compared to initial conditions from state-of-the-art data assimilation in ERA5. This enables the application of the method to real-world applications, such as creating reanalysis datasets with autoregressive data assimilation.

\end{abstract}

\section{Introduction}
\begin{figure}[h!]
    \centering
    \includegraphics[width=\linewidth]{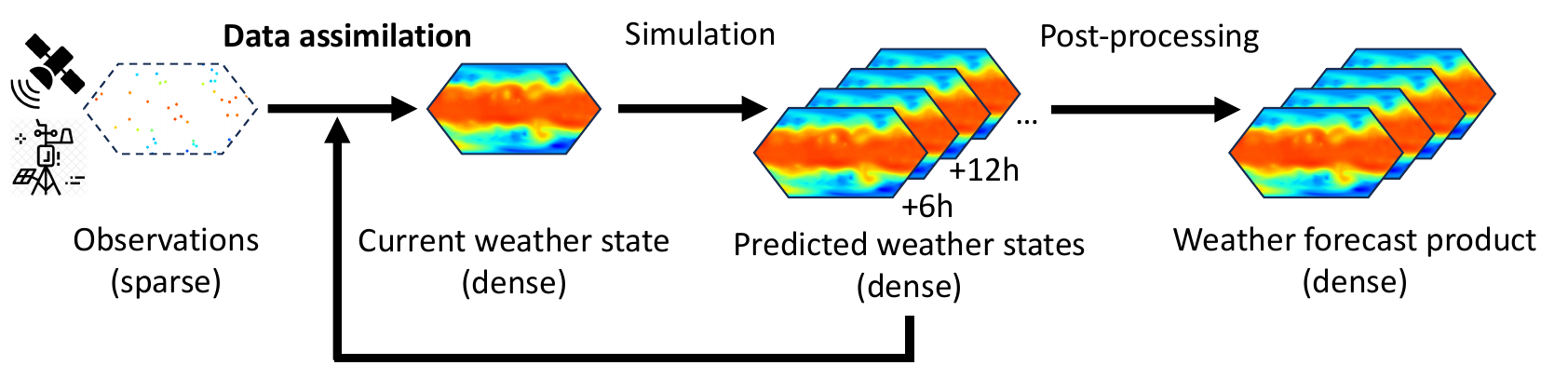}
    \caption{Diagram of a numerical weather forecasting pipeline. It consists of data assimilation, simulation and post-processing. Data assimilation produces gridded values from sparse observations and predicted gridded values from previous time steps. Simulation takes in gridded values and produces predictions in gridded values at future time steps. Post-processing improves prediction so that it is closer to future observations.}
    \label{fig:diagram_pipeline}
\end{figure}
Weather prediction plays an important role in our daily life. Due to uncertainties in the prediction process,  predicted weather will inevitably deviate from actual weather. It is necessary to ``pull-back'' the predicted weather state via \emph{data assimilation}, to align it with observations from weather stations and satellites. Practically, as \autoref{fig:diagram_pipeline} shows, data assimilation generates initial conditions for weather simulation models using predicted weather states and sparse observations from various locations.
The quality of these weather simulation models depends heavily on data assimilation, as errors in initial conditions are one of the main sources of forecast error~\citep{bonavita2016ecmwfDA}. Additionally, data assimilation is employed in creating reanalysis datasets, which contain reconstructed historical weather variables as gridded fields. These reanalysis datasets play a central role in weather and climate research~\citep{ipcc21PB, hersbach2020era5}, and are essential for the training of ML weather forecasting models~\citep{pathak2022fourcastnet, bi2023panguweather, lam2023graphcast}.

\begin{figure*}[h!]
    \centering
    \includegraphics[width=0.75\linewidth]{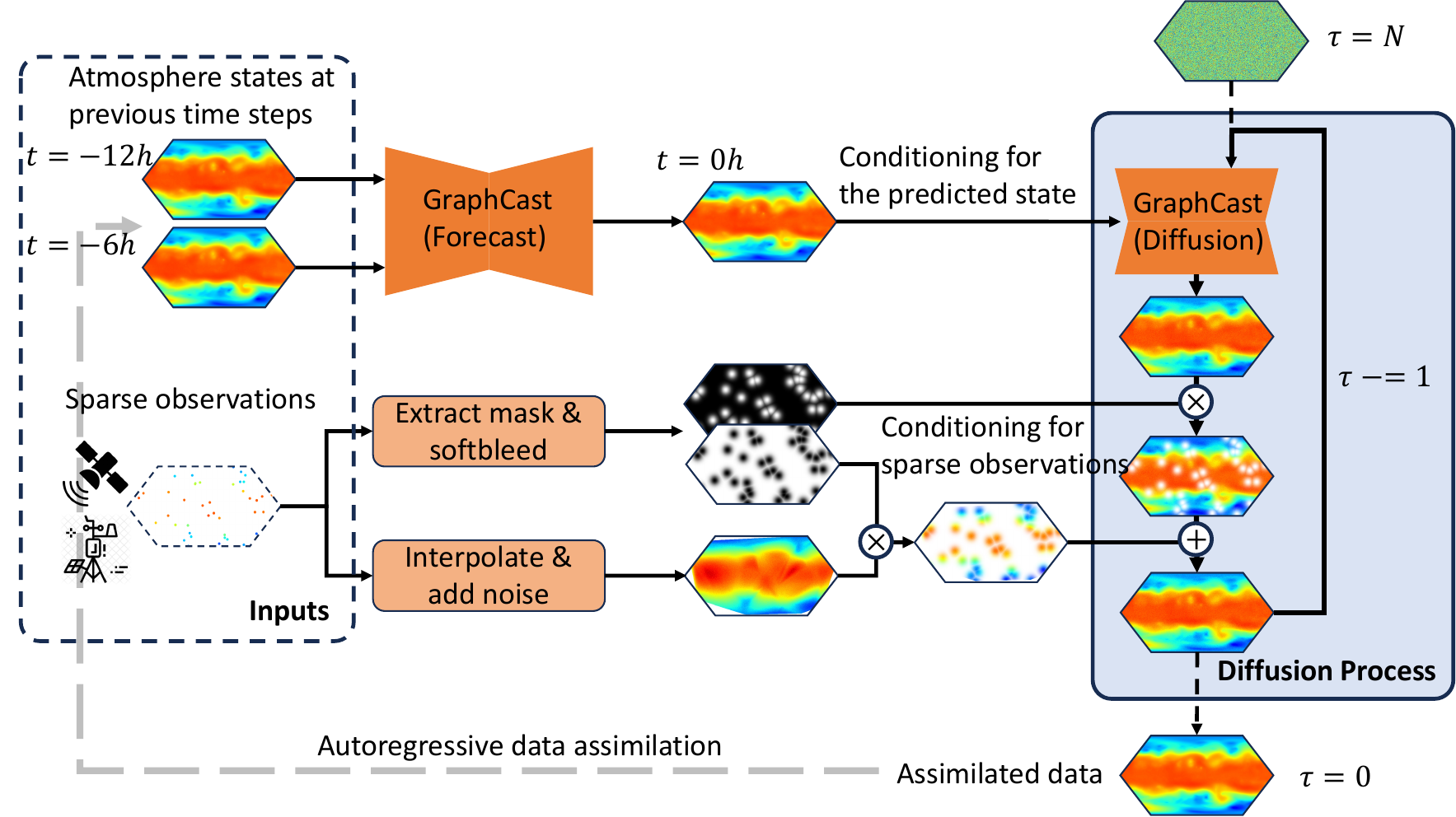}
    \caption{Architecture of the diffusion-based data assimilation method. We take advantage of the input and output shape of the pretrained GraphCast model, which takes the state of the atmosphere at two time steps as input. In each iteration of the denoising diffusion process, the adapted GraphCast model takes the predicted state and the assimilated state with noise, and further denoises the assimilated state. To enforce the observation values at inference time, The denoised state is merged with interpolated observations using a soft mask created by softbleeding the hard mask derived from the original observations.}
    \label{fig:diagram_model}
\end{figure*}

Various data assimilation methods have been developed and employed to address different characteristics of observation data and system dynamics. Among these methods, variational data assimilation and the ensemble Kalman filter are the most widely used methods in operational data assimilation~\citep{bannister2017VarEnKFReview}. The variational method solves an optimization problem by minimizing a cost function that measures the discrepancy between simulations and observations. It requires multiple iterations in which linearized observation and evolution functions are evaluated to compute the gradient of the cost function. The linearized and original versions of observation and evolution functions are often implemented separately. This adds overhead because the linearized functions have similar code complexity as the original functions, and a lot of effort has to be invested in maintaining the consistency of the original and linearized code.
The ensemble Kalman filter method updates the state estimation according to the covariance matrix calculated from ensemble simulations. Both approaches are computationally intensive as one requires multiple optimization iterations and the other requires multiple ensemble simulations.

Data assimilation tools are becoming a bottleneck in the simulation pipeline. While traditional data assimilation methods are sufficiently competent in operational weather forecasts~\citep{bonavita2016ecmwfDA}, but their high computational and development costs restrict broader adoptions, making them tightly coupled with a specific numerical weather forecasting code. This restriction becomes more evident since the explosion of ML weather forecasting models~\citep{pathak2022fourcastnet,lam2023graphcast, bi2023panguweather}. Those models claim to be a replacement of the traditional models by achieving competitive or even superior accuracy compared to the best operational model while being orders of magnitude faster. Ironically, they cannot independently make forecasts as they are all trained and evaluated on the ERA5 dataset~\citep{hersbach2020era5} which is produced by the traditional data assimilation method together with the numerical forecast model. 


In a probabilistic view, data assimilation can be formulated as sampling from a probability distribution of atmosphere states conditioned on observations and predicted states~\citep{law2015DAbook, evensen2022DAfundamentals}. Capable of solving this conditional sampling problem, denoising diffusion models~\citep{ho2020ddpm} naturally become a tentative choice for data assimilation. Moreover, the blooming community of diffusion models provides an arsenal of techniques for enforcing conditions of different kinds. In particular, conditioning techniques for in-painting~\citep{lugmayr2022repaint, song2020score} and super-resolution~\citep{saharia2022imagen,chung2022dps} are of special interest, because they are similar to conditioning of observations and predicted states respectively. This denoising diffusion model approach has been applied to relatively small-scale data assimilation problems~\citep{rozet2023scoreDA, rozet2023scoreDAGeostrophic, finn2023representationDDM, andry2023diffusionDA}, but none can assimilate data with a resolution comparable with the ERA5 dataset (0.25$^\circ$ horizontal resolution), limiting their use with ML forecast models. Similarly, the denoising diffusion techniques have been applied in weather forecasts~\citep{price2023gencast} and post-processing~\citep{mardani2023diffNV, mardani2024residualNV2, li2023seeds} with either limited resolution or limited region.
In this work, we propose a new approach to data assimilation based on the denoising diffusion model with a focus on weather and climate applications. We are able to scale to 0.25$^\circ$ horizontal resolution with 13 vertical levels by utilizing the ML forecast model GraphCast~\citep{lam2023graphcast} as the backbone of the diffusion model. As we do not have real-world observations available from the ERA5 dataset, we are using grid columns of the ERA5 reanalysis dataset as proxies for observations.
During training, the diffusion model is conditioned with the predicted state, i.e., the atmospheric state produced by the forecast model from earlier initial conditions. During inference, we further condition the model with sparse column observations following Repaint~\citep{lugmayr2022repaint}. In addition, we use a soft mask and interpolated observations to strengthen the conditioning utilizing the continuity of atmospheric variables. The resulting assimilated data can converge to the ground truth field when the number of simulated observation columns increases. More importantly, the assimilated data can be used as inputs to the forecast model at a loss of lead time not exceeding 24 hours. We also test the autoregressive data assimilation to generate a reanalysis dataset given a series of observations and an initial predicted field. 

Our key contributions are:
\begin{enumerate}
 \item{We demonstrate a novel ML data assimilation method capable of assimilating high resolution data. The assimilated data is ready for weather forecasting applications.}
 \item{We create data assimilation cycles by combining our method with an ML weather forecasting model. The resulting reanalysis dataset is consistent with the given observations.}
 \item{We build our method using a neural network backbone from a pretrained ML forecast model. It is easy to upgrade the backbone with any state-of-the-art model due to the flexibility of our method.}
\end{enumerate}


\section{Method}
\subsection{Problem Formulation}\label{sec:prob_formulation}
The goal of data assimilation is to reconstruct atmospheric variables on a fixed grid with $n$ grid points $\mathbf{x}_i\in \mathbb{R}^n$ at physical time step $i$ given $m$ measurements $\mathbf{y}_i=f(\mathbf{x}^*_i), \mathbf{y}_i \in \mathbb{R}^m, \mathbf{x}_i^* \in \mathbb{R}^n$ where $\mathbf{x}^*_i$ is the ground truth of atmospheric variables on grid points at time step $i$. In addition, estimated values on grid points $\hat{\mathbf{x}}_i = \mathcal{F}(\mathbf{x}_{i-1})$ produced by the forecast model $\mathcal{F}: \mathbb{R}^n \rightarrow \mathbb{R}^n$ are also provided as one of the inputs. In a probabilistic view, data assimilation samples from a conditional distribution $p(\mathbf{x}_i|\hat{\mathbf{x}}_i, \mathbf{y}_i)$ which minimizes the discrepancy between $\mathbf{x}_i$ and $\mathbf{x}^*_i$. To simplify the problem, $f$ is limited to a sparse linear observation operator $\mathbf{y}_i = f(\mathbf{x}^*) = \mathbf{Ax}^*$ where $\mathbf{A}$ is a sparse matrix with only one nonzero value in each row. In real-world cases, this simplification applies to point observations such as temperature, pressure, and wind speed measurements at weather stations and balloons. 

\subsection{Denoising Diffusion Probabilistic Model}
The denoising diffusion probabilistic model (DDPM) is a generative model capable of sampling from the probabilistic distribution defined by the training data~\citep{ho2020ddpm}. It is trained to approximate the reverse of the diffusion process where noise is added to a state vector $\mathbf{x}^0$ during $N$ diffusion steps, resulting in an isotropic Gaussian noise vector $\mathbf{x}^N \sim \mathcal{N}(\mathbf{0},\mathbf{I})$. We denote the state vector at physical time step $i$ and diffusion step $j$ with $\mathbf{x}^j_i$. We write $\mathbf{x}^j$ whenever the statement is independent of the physical time step. Note that the diffusion step $j$ and the physical time step $i$ are completely independent of each other. 

For each diffusion step $j$, the diffusion process can be seen as sampling from a Gaussian distribution with a mean of $\sqrt{1-\beta_j}\mathbf{x}^{j-1}$ and covariance matrix of $\beta_j \mathbf{I}$:
\begin{equation}
    p(\mathbf{x}^j | \mathbf{x}^{j-1}) = \mathcal{N}(\sqrt{1-\beta_j}\mathbf{x}^{j-1}, \beta_j \mathbf{I}) \label{eq:forward}
\end{equation}
where $\beta_j > 0$ is the variance schedule. 

A denoising diffusion model $\mathbf{\mu}_\theta$ is used to predict the mean of $\mathbf{x}^{j-1}$ given $\mathbf{x}^j$ and $j$ with the following parameterization:
\begin{equation}
    \mu_\theta(\mathbf{x}^j, j) = \frac{1}{\sqrt{1-\beta_j}}\left( \mathbf{x}^j - \frac{\beta_j}{\sqrt{1-\bar{\alpha}_j}}\epsilon_\theta(\mathbf{x}^j, j) \right) \label{eq:ddpm_mu}
\end{equation}
 where $\theta$ are the trainable parameters, $ \bar{\alpha}_j = \prod_{s=1}^j{(1-\beta_s)}$. Then, $\mathbf{x}^{j-1}$ can be sampled from $p(\mathbf{x}^{j-1} | \mathbf{x}^j) =\mathcal{N}(\mu_\theta(\mathbf{x}^j, j), \frac{1-\bar{\alpha}_{j-1}}{1- \bar{\alpha}_j}\beta_t \mathbf{I})$ to reverse the diffusion process. Applying this procedure $N$ times from $\mathbf{x}^N \sim \mathcal{N}(\mathbf{0},\mathbf{I})$, we can generate $\mathbf{x}^0$ which follows a similar distribution as the empirical distribution $\mathcal{D}$ of the training data.
 
During training, we minimize the following loss function:
\begin{equation*}
    L(\theta) = \mathbb{E}_{j\sim U[1, N], \mathbf{x}^{0} \sim \mathcal{D}, \epsilon \sim \mathcal{N}(\mathbf{0}, \mathbf{I})}\left[ \| \epsilon - \epsilon_\theta(\mathbf{x}^j, j) \|^2 \right] .
\end{equation*}
Note that $\mathbf{x}^j$ can be expressed in closed form as $\mathbf{x}^j =~\sqrt{\bar{\alpha}_j}\mathbf{x}^{0} + \sqrt{(1 - \bar{\alpha}_j)} \epsilon$, because the diffusion process applies independent Gaussian noise at each step, and thus $p(\mathbf{x}^j | \mathbf{x}^{0}) = \mathcal{N}(\sqrt{\bar{\alpha}_j}\mathbf{x}^{0}, (1 - \bar{\alpha}_j) \mathbf{I})$.


With the unconditional denoising diffusion model above, we add the conditioning of predicted state $\hat{\mathbf{x}}$ and observation $\mathbf{y}$ separately according to their characteristics.




\subsection{Conditioning for Predicted State}\label{sec:cond_pred}
Utilizing the fact that the predicted state $\hat{\mathbf{x}}$ has the same shape as the diffused state $\mathbf{x}^j$, it is convenient to add the predicted state as an additional input for the (reparameterized) diffusion model denoted as $\epsilon_\theta(\mathbf{x}^j, \hat{\mathbf{x}}, j)$~\citep{saharia2022imagen}. The underlying neural network can concatenate $\mathbf{x}^j$ and $\hat{\mathbf{x}}$ in their feature channels without changing its architecture as shown in \autoref{fig:diagram_model}.

Since $\epsilon_\theta$ (and $\mu_\theta$) is dependent on the predicted state $\hat{\mathbf{x}}$, the reverse diffusion process becomes sampling from $p(\mathbf{x}^{j-1} | \mathbf{x}^j, \hat{\mathbf{x}}) =~\mathcal{N}(\mu_\theta(\mathbf{x}^j, 
\hat{\mathbf{x}}, j), \frac{1-\bar{\alpha}_{j-1}}{1- \bar{\alpha}_j}\beta_j \mathbf{I})$. Therefore, the generated $\mathbf{x}^0$ is sampled from the conditional distribution
$p(\mathbf{x}^0 | \hat{\mathbf{x}}) =~p(\mathbf{x}^N)\prod_{j=1}^N{p(\mathbf{x}^{j-1}| \mathbf{x}^j, \hat{\mathbf{x}})}$.

This method requires sampling pairs of $\mathbf{x}^{0}$ and $\hat{\mathbf{x}}$ during training. This can be realized by sampling two states $\mathbf{x}_{i-1}^*, \mathbf{x}_i^*$ with consecutive physical time steps from training data, and then apply the forecast model to get the predicted state at physical time $i$: $\hat{\mathbf{x}}_i = \mathcal{F}(\mathbf{x}_{i-1}^*)$. Afterwards, the sampled ground truth state $\mathbf{x}_i^*$ is diffused with a random number $j\in [1, N]$ steps. This process is implemented with the closed form formula  $\mathbf{x}^j_i =~\sqrt{\bar{\alpha}_j}\mathbf{x}^{*}_i +~\sqrt{(1 - \bar{\alpha}_j)} \epsilon, \enspace \epsilon \sim~\mathcal{N}(\mathbf{0}, \mathbf{I})$. Finally, we take a gradient descent step based on $\nabla_\theta \| \epsilon -~\epsilon_\theta(\mathbf{x}^j_i, \hat{\mathbf{x}}_i, j) \|^2$ to train the diffusion model.

With the conditioning on the predicted state alone, we can already perform denoising diffusion steps using the trained model $\mu_\theta$, resulting in $\mathbf{x}^0 \sim p(\mathbf{x}^0 | \hat{\mathbf{x}})$. This means we corrected, or in other words, ``post-processed'' the predictions $\hat{\mathbf{x}}$ to be closer to the ground truth state $\mathbf{x}^*$.

\subsection{Conditioning for Sparse Observations}
The conditioning for sparse observations poses a different challenge than the conditioning on the predicted state. The sparse observations $\mathbf{y}$ have a variable length $m$ as opposed to a fixed length $n$, and the data assimilation results are invariant to the permutation of the $m$ elements in $\mathbf{y}$. This requires a dedicated design in the neural network if we want to directly condition $\mu_\theta$ with $\mathbf{y}$ as before. Even if we find a solution to implement that, the trained diffusion model will have a generalization problem because the possible input space spanned by $\mathbf{x}^0$, $\hat{\mathbf{x}}$, and $\mathbf{y}$ is too large and hard to thoroughly sample during training.

To avoid those issues, we follow inpainting techniques~\citep{lugmayr2022repaint, song2020score} to enforce the conditioning on the sparse observations at inference time. Let us start with a simple approach first. It creates a hard mask $\mathbf{m}_h\in \{0, 1\}^n$ to indicate which grid columns of the atmospheric state are observed, where a 1 means the associated value in $\mathbf{x}^*$ is observed in $\mathbf{y}$ and vice versa. The mask equals the sum of the columns of the observation matrix $\mathbf{A}\in \mathbb{R}^{n\times m}$ where $\mathbf{y} = \mathbf{Ax}^*$. During the inference of the diffusion model, a state vector with white Gaussian noise is created and gradually denoised by the denoising diffusion model. However, with $\mathbf{y}$ present, we have better knowledge of the observed locations in the state vector: values can be produced by a forward diffusion process from the observation data. We can use the mask $\mathbf{m}_h$ to treat the two parts separately and combine them in each iteration of the denoising process (\autoref{fig:diagram_model}).
\begin{align*}
    \mathbf{x}^{j-1}_{\text{known}} & \sim \mathcal{N}(\sqrt{\bar{\alpha}^{j-1}}\mathbf{x}^*, (1-\bar{\alpha}^{j-1})\mathbf{I}) \\
    \mathbf{x}^{j-1}_{\text{unknown}} & \sim \mathcal{N}(\mu_\theta(\mathbf{x}^j, j), \frac{1-\bar{\alpha}_{j-1}}{1- \bar{\alpha}_j}\beta_t \mathbf{I}) \\
    \mathbf{x}^{j-1} & = \mathbf{m}_h \odot  \mathbf{x}^{j-1}_{\text{known}} + (1 - \mathbf{m}_h) \odot \mathbf{x}^{j-1}_{\text{unknown}}
\end{align*}
Here, $\odot$ denotes point-wise multiplication used to filter observed and non-observed values. Although this method can guide diffusion results to the observed values, it performs poorly in practice because only the values at the observation locations are forced to the given values while other values remain unchanged as in the unconditional scenario. This likely relates to the encoding-process-decoding architecture commonly used in diffusion models: the encoding and decoding layers employ pooling to downscale and upscale in spatial dimensions. While it helps to condense information and reduce calculations, it also smears out local details. As a result, the added conditioning information is often lost during this process.

\begin{figure}[t!]
    \vskip -0.05in
    \centering
    \includegraphics[width=0.8\linewidth]{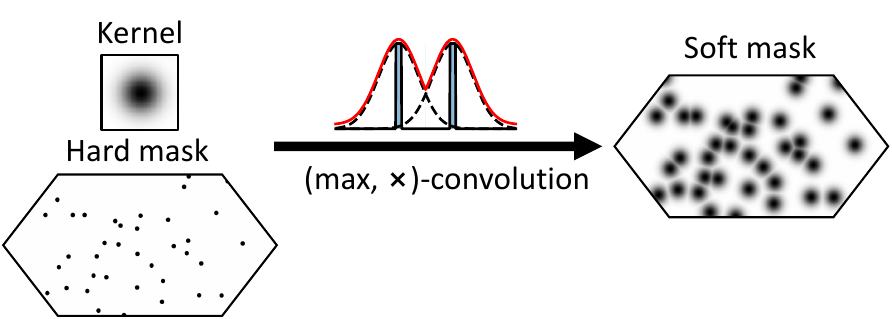}
    \caption{Creating a soft mask from a hard mask using softbleed. Softbleed performs a ($\max$,$\times$)-convolution over the Gaussian kernel and the hard mask. }
    \label{fig:diagram_softbleed}
\end{figure}

We tackle this issue using a ``soft mixing" instead of a hard one. In the soft mixing, we replace the hard mask $\mathbf{m}_h$ with a soft mask $\mathbf{m}_s$ by applying softbleeding to $\mathbf{m}_h$ (\autoref{fig:diagram_softbleed}) with standard deviation $\sigma_G$ (for the Gaussian kernel), and diameter $d$. The softbleed process mimics the Gaussian blurring but replaces the convolution with the ($\max$,$\times$)-convolution to ensure the 1-valued points in the hard mask remain the same in the soft mask.
The support region of $\mathbf{m}_s$ (where its values are larger than 0) is larger than the support region of the observed values. As the atmospheric variables are relatively continuous over space, we interpolate the observed values to fill the support region of $\mathbf{m}_s$. 
There is abundant flexibility in choosing interpolation algorithms. We use the universal kriging interpolation algorithm provided in the \textsc{pyinterp} package. In summary, the inference iteration becomes:
\begin{align}
\begin{split}
    \mathbf{m}_s & = \text{Softbleed}(\mathbf{m}_h, \sigma_G, d) \\
    {\mathbf{x}^*}' & = \text{Interpolate}(\mathbf{m}_h \odot \mathbf{x}^*, d) \\
    \mathbf{x}^{j-1}_{\text{known}} & \sim \mathcal{N}(\sqrt{\bar{\alpha}^{j-1}}{\mathbf{x}^*}', (1-\bar{\alpha}^{j-1})\mathbf{I}) \\
    \mathbf{x}^{j-1}_{\text{unknown}} & \sim \mathcal{N}(\mu_\theta(\mathbf{x}^j, j), \frac{1-\bar{\alpha}_{j-1}}{1- \bar{\alpha}_j}\beta_t \mathbf{I}) \\
    \mathbf{x}^{j-1} & = \mathbf{m}_s \odot  \mathbf{x}^{j-1}_{\text{known}} + (1 - \mathbf{m}_s) \odot \mathbf{x}^{j-1}_{\text{unknown}} . 
\end{split}\label{eq:backward_cond2}
\end{align}

In addition, we also applied the resampling technique from~\citep{lugmayr2022repaint} to further reduce the inconsistency between the known part and unknown part. For each denoising iteration, the resampling technique repeats the iteration $U$ times by adding noise to $\mathbf{x}^{j-1}$ to create $\mathbf{x}^{j}$ following \autoref{eq:forward} and repeating the denoising step \autoref{eq:backward_cond2}. The overall algorithm of applying the denoising diffusion model for data assimilation is presented in Algorithm \autoref{alg:inference}.


\begin{algorithm}
  \caption{Data assimilation (inference)}\label{alg:inference}
  \begin{algorithmic}
  \REQUIRE predicted state $\hat{\mathbf{x}}$, hard mask of observations $\mathbf{m}_h$, observation values at grid points $\mathbf{m}_h\odot\mathbf{x}^*$ (created from observation vector $\mathbf{y}$, and observation operator $\mathbf{A}$), 
  covariance schedule $\beta_j, j=1,\cdots,N$, Gaussian kernel standard deviation $\sigma_G$, Gaussian kernel diameter $d$, scaling factor $\mathbf{s}$ for de-normalization
  \ENSURE $\mathbf{x} \sim p(\mathbf{x} | \hat{\mathbf{x}}, \mathbf{y})$
  \STATE $\mathbf{m}_s = \text{Softbleed}(\mathbf{m}_h, \sigma_G, d)$
  \STATE ${\mathbf{x}^*}' = \text{Interpolate}(\mathbf{m}_h \odot \mathbf{x}^*, d)$
  \STATE $\mathbf{x}^N \sim \mathcal{N}(\mathbf{0},\mathbf{I})$
  \FOR{j in $N, \cdots, 1$}
    \HEADER{Conditioning for predicted state}
      \STATE $\bar{\alpha}_j = \prod_{s=1}^j{(1-\beta_s)}$
      \STATE $\bar{\mathbf{x}}^{j-1}_{\text{unknown}} = \frac{1}{\sqrt{1-\beta_j}}\left( \mathbf{x}^j - \frac{\beta_j}{\sqrt{1-\bar{\alpha}_j}}\epsilon_\theta(\mathbf{x}^j, \hat{\mathbf{x}}, j) \right)$
      \STATE $\mathbf{x}^{j-1}_{\text{unknown}} \sim \mathcal{N}(\bar{\mathbf{x}}^{j-1}_{\text{unknown}}, \frac{1-\bar{\alpha}_{j-1}}{1- \bar{\alpha}_j}\beta_j \mathbf{I})$
    \ENDHEADER
    \HEADER{Conditioning for sparse observations}
      \STATE $\mathbf{x}^{j-1}_{\text{known}} \sim \mathcal{N}(\sqrt{\bar{\alpha}^{j-1}}{\mathbf{x}^*}', (1-\bar{\alpha}^{j-1})\mathbf{I})$
      \STATE $\mathbf{x}^{j-1} = \mathbf{m}_s \odot  \mathbf{x}^{j-1}_{\text{known}} + (1 - \mathbf{m}_s) \odot \mathbf{x}^{j-1}_{\text{unknown}}$
    \ENDHEADER
  \ENDFOR
  \STATE $\mathbf{x} = \hat{\mathbf{x}} + \mathbf{s}\odot\mathbf{x}^0$ (apply skip connection)
\end{algorithmic}
\end{algorithm}


\subsection{Selection of Diffusion Model Structure}
Our method provides a lot of flexibility in the selection of the neural network structure for $\epsilon_\theta(\mathbf{x}^j_i, \hat{\mathbf{x}}_i, j): \mathbb{R}^n\times\mathbb{R}^n\times\mathbb{N}^+\rightarrow\mathbb{R}^n$ as any neural network that matches the function signature will work. However, $n$ can be tens of millions in practice, rendering many neural network architectures infeasible due to resource constraints. A neural network must utilize the spatial information in the state vector $\mathbf{x}$ to learn efficiently.

Instead of creating a new architecture, we adapt a proven ML weather forecasting model architecture. Such models have a similar signature $\mathbb{R}^{c\cdot n}\rightarrow\mathbb{R}^n, c\in \mathbb{N}^+$ as the diffusion model. Due to this similarity, neural networks that perform well in forecasting should also do well in the data assimilation tasks, and it is likely to take fewer training steps using pretrained weights of the forecast model when training the diffusion model. Moreover, we can easily update the backbone of the diffusion model with the state-of-the-art weather forecasting model. 



\section{Experiments}

\subsection{Implementation}
We demonstrate our method in a real-world scenario containing 6 pressure-level variables (temperature, geopotential, u-wind, v-wind, vertical velocity, specific humidity) and 4 surface variables (2m temperature, 10m u-wind, 10 v-wind, mean sea level pressure), with a horizontal resolution of 0.25$^\circ$ and 13 vertical levels (50hPa, 100hPa, 150hPa, 200hPa, 250hPa, 300hPa, 400hPa, 500hPa, 600hPa, 700hPa, 850hPa, 925hPahPa, 1000hPa). This matches the resolution of the WeatherBench2 dataset~\citep{rasp2023weatherbench2} used by state-of-the-art ML weather forecasting models. We use the GNN-based GraphCast model as the backbone of the diffusion model because the pretrained model takes in states at two consecutive time steps $\mathbf{x}_{i-1}, \mathbf{x}_i$ to predict $\mathbf{x}_{i+1}$. It takes much less effort than other forecast models to re-purpose it to $\epsilon_\theta(\mathbf{x}^j_i, \hat{\mathbf{x}}_i, j)$ given that there is one-to-one matching between $(\mathbf{x}_{i-1}, \mathbf{x}_{i}, \mathbf{x}_{i+1})$ and $(\hat{\mathbf{x}}_i, \mathbf{x}^j_i. \mathbf{x}^{j-1}_i)$. As is determined by the pre-trained GraphCast model, the input size $n$ is set to $(6\times 13 + 4)\times 721 \times 1440$.

The diffusion model is implemented with the \textsc{jax} library~\citep{jax2018jax}, \textsc{diffusers} library~\citep{Diffusers}, and the official implementation of GraphCast. We use the AdamW optimizer~\citep{loshchilov2018adamw} with a warm-up cosine annealing learning rate schedule that starts from $10^{-5}$, peaks at $10^{-4}$ after 1/6 of total training steps, and ends at $3\times10^{-6}$. We perform data-parallel training on 48 NVIDIA A100 GPUs with a (global) batch size of 48 for 20 epochs. Gradient checkpoints are added in the GraphCast model to reduce the GPU memory footprint. The training takes approximately 2 days. All the inference is performed on a single node with one A100 GPU, which produces an assimilated state in around 15 minutes.

\subsection{Training Data}
We use the WeatherBench2 dataset as the first part of the training data. The dataset contains values for our target atmospheric variables from 1979 to 2016 with a time interval of 6 hours extracted from the ERA5 reanalysis dataset. The training process uses data from 1979 to 2015. Validation uses data in 2016 and testing uses data in 2022. The second part of the training data is generated by running 48-hour GraphCast predictions (with 8 time steps) using the data from ERA5 as initial conditions. Then, the two parts are paired up according to their physical time. Before feeding it to the model, the input data is normalized using the vertical-level-wise means and standard deviations provided by the pre-trained GraphCast model. The output of the model is de-normalized using the same means and standard deviations. Since the predicted state $\hat{\mathbf{x}}$ is close to the ground truth $\mathbf{x}^*$, we add a skip connection from $\hat{\mathbf{x}}$ to the de-normalized model output. For the autoregressive data assimilation experiment, we train an additional data assimilation model dedicated for 6-hour forecasts.


\subsection{Treatment of Conditioning for Sparse Observations}
Acknowledging the multidimensional nature of the state vector and that most meteorological observations are co-located horizontally (longitude and latitude), we opt for a simplified setting in the conditioning of sparse observations. In this scenario, the observational data is $m$ sampled columns of the ground truth state vector with $6\times 13 + 4$ values in each column: $\mathbf{y}\in \mathbb{R}^{(6\times 13 +4)\times m}$. We set Gaussian kernel standard deviation $\sigma_G$ to 2.5 according to the ablation study and kernel diameter $d$ to $4\sigma_G+1$ (using number of grid cells as the unit). As kernels are applied to observations at different latitudes, $\sigma_G$ is scaled with cosine of latitude according to the sphere geometry. The mask is simplified to a 2D mask $\mathbf{m}_s \in \mathbb{R}^{721\times 1440}$ which is broadcast to other dimensions when doing point-wise multiplication with the state vector in Equation \autoref{eq:backward_cond2}. Interpolation is also applied independently over 2D horizontal slices for each variable and level. We use the universal kriging interpolation method provided by the \textsc{pyinterp} package to properly interpolate unstructured data points on the sphere. The data points are subtracted by climatology values from WeatherBench2 to conform to the prerequisites of kriging interpolation.

\subsection{Experiment Settings and Results}
\begin{figure}[h!]
    \centering
    \includegraphics[width=\linewidth]{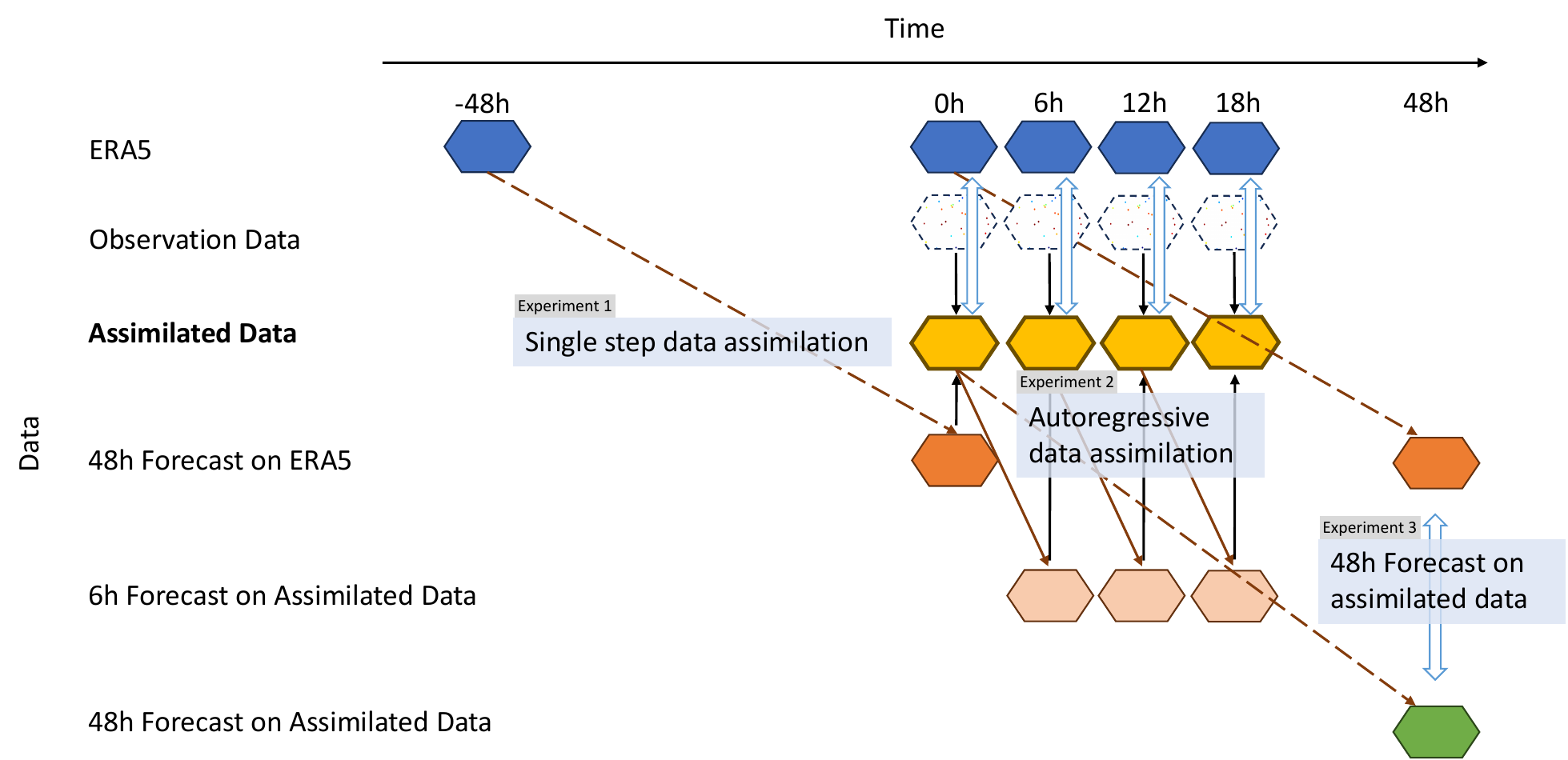}
    \caption{Overview of the experiment settings. Single-step data assimilation takes in observations and a 48h forecast, and outputs assimilated data at 0h. Autoregressive data assimilation combines a data assimilation model and a 6h prediction model to produce assimilated data every 6h autoregressively. It is also of interest to perform 48h forecasts on single-step assimilated data.  Hexagons represent atmosphere states, black arrows represent data assimilation, brown solid arrows represent 6-hour prediction, brown dashed arrows represent 48-hour prediction, hexagons with dashed edges and sparse points represent sparse observations, and wide arrows point out targets and references to compare in each experiment. 
    }
    \label{fig:exp_diagram}
\end{figure}
We demonstrate the effectiveness of our method by performing real-world inspired experiments of increasing sophistication (\autoref{fig:exp_diagram}). 
First, in the \emph{single step} experiment, we perform data assimilation on 48-hour forecast and observation data, then directly compare the assimilated data with the ground truth. Second, we evaluate our method with \mbox{\emph{autoregressive}} data assimilation, where a 6h prediction and data assimilation cycle is run repeatedly. We design this experiment to test whether the assimilated data will deviate from the ground truth data after several autoregressive iterations, which is crucial in real-world applications. Lastly, we designed an experiment to compare the 48-hour forecast based on assimilated data and ERA5 data and evaluate the effect of the data assimilation method on the \emph{forecast} skill. For single-step data assimilation and forecast on single-step data assimilation, we run 16 parallel experiments with inputs from different time steps and different random seeds. For each experiment, we calculate the assimilation errors against the ERA5 dataset in root mean square errors (RMSEs) and then present the averaged RMSEs in the results. For the autoregressive data assimilation, we run each experiment once due to the limitations of computation resources.
In all the experiments, observations are simulated by taking random columns from the ERA5 dataset considered as the ground truth. We vary the number of observed columns $m$ in the experiments to test the convergence property of our data assimilation method.


\begin{figure}[h!]
    \centering
    \includegraphics[width=\linewidth]{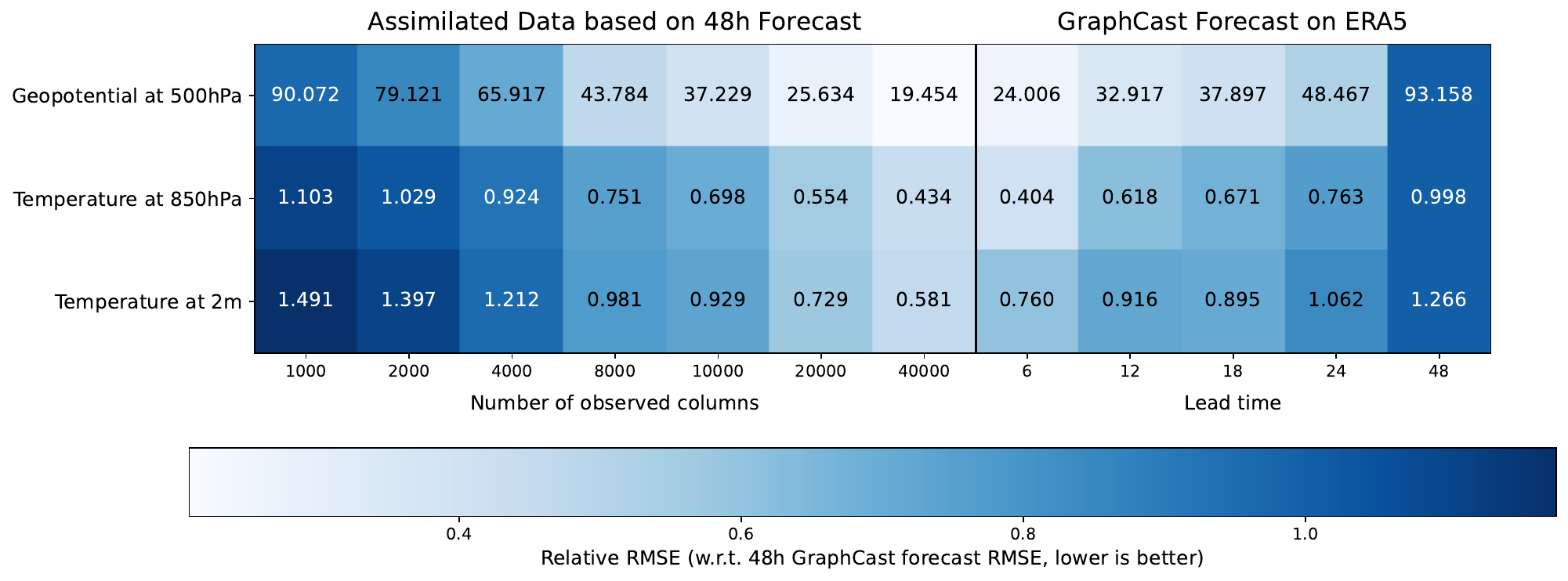}
    \caption{Root mean square errors (RMSEs, shown by the numbers in the cell) of geopotential at 500hPa, temperature at 850hPa, and temperature at 2m from the single-step assimilated data, and from 6-hour to 48-hour GraphCast forecasts. The errors are calculated against the ERA5 data. The cells are color-coded with the RMSEs relative to the 48-hour forecast errors.}
    \label{fig:exp1_score}
\end{figure}



\paragraph{Single-step Data Assimilation} In this experiment, we perform data assimilation from a 48-hour prediction using GraphCast and columns of observations ranging from 1,000 to 40,000, then calculate the RMSE errors between the assimilated data and the ERA5 data. Finally, we compare the errors against GraphCast forecast errors with 6-hour to 48-hour lead times. 

The result is presented in \autoref{fig:exp1_score} where we pick three representative variables closely related to forecast skills~\citep{ashkboos2022ens10, rasp2020weatherbench} including geopotential at 500hPa (z500), temperature at 850hPa (t850), and temperature at 2m (t2m) as the target variables. More results can be found in \autoref{sec:scoreboard1}. Generally, assimilation errors decrease as number of observed columns increases. With 4,000 observed columns ($< 0.4\%$ total data), the assimilated data achieves lower RMSEs than the input 48-hour forecast for all three variables. The errors further decrease as the number of observed columns increases. On the other side of the spectra, with 40,000 observed columns ($< 3.9\%$ total columns), the RMSEs of z500 and t2m are lower than the 6-hour forecast error. 

\begin{figure}[h!]
    \centering
    \includegraphics[width=\linewidth]{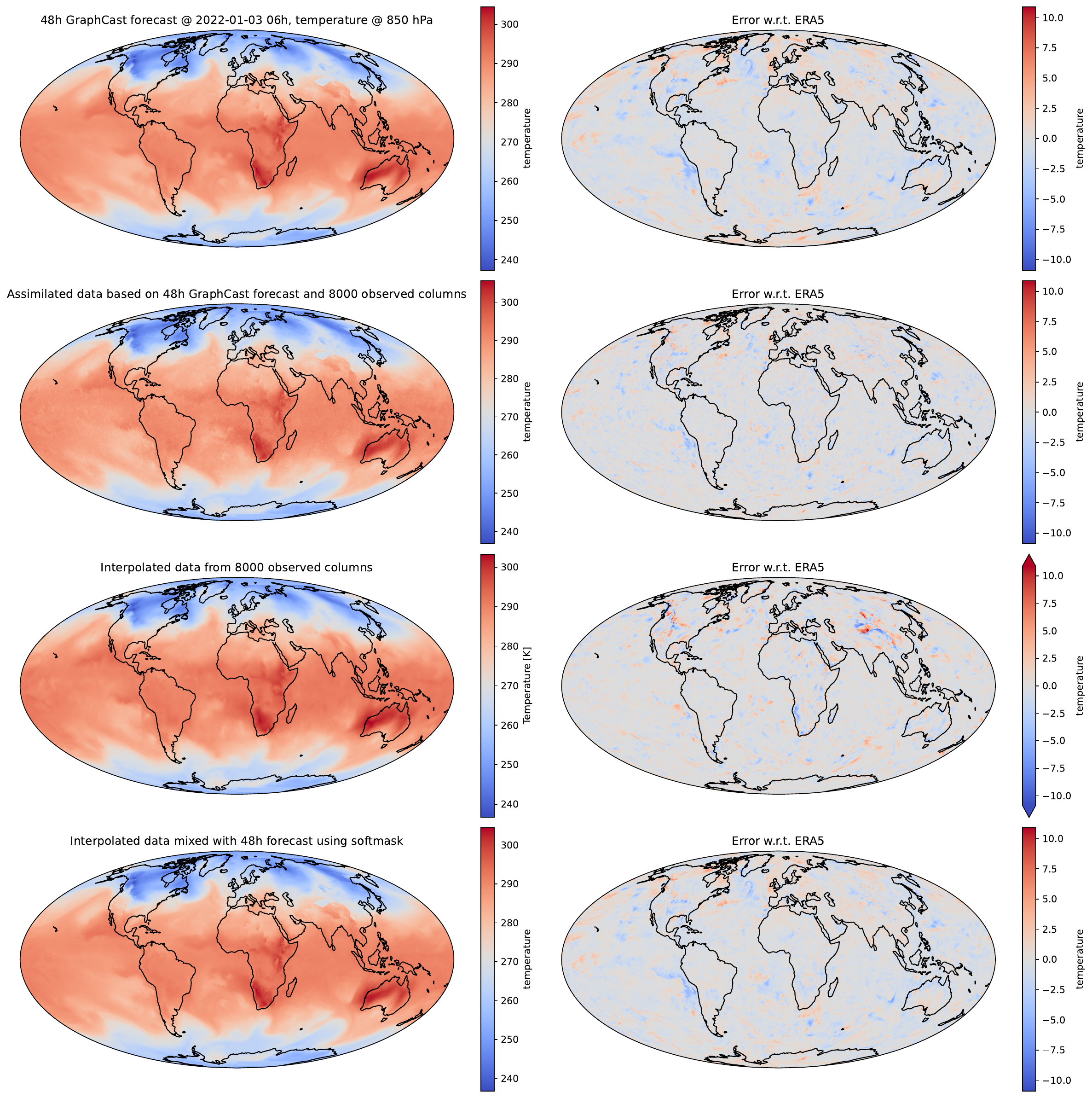}
    \caption{Plots of assimilated temperature at 850hPa and errors against ERA5 compared with 48h GraphCast forecast, interpolated observations, and mixture of 48h forecast and interpolated observation according to the soft mask. The assimilated data is better in accuracy than both inputs. It also introduces more fine grained details than 48h forecast. }
    \label{fig:casestudy}
\end{figure}

As a case study, we present the assimilated t850 with 8000 observed columns in \autoref{fig:casestudy}. The assimilated data have lower errors than the two inputs of data assimilation (48h forecast and interpolated observations). It is also better than simply mixing forecast with interpolated observations. In addition, the assimilated data introduce fine grained details which helps in recreating high-frequency spectral component lost during autoregressive forecast.

\begin{figure}[h!]
    \centering
    \includegraphics[width=0.85\linewidth]{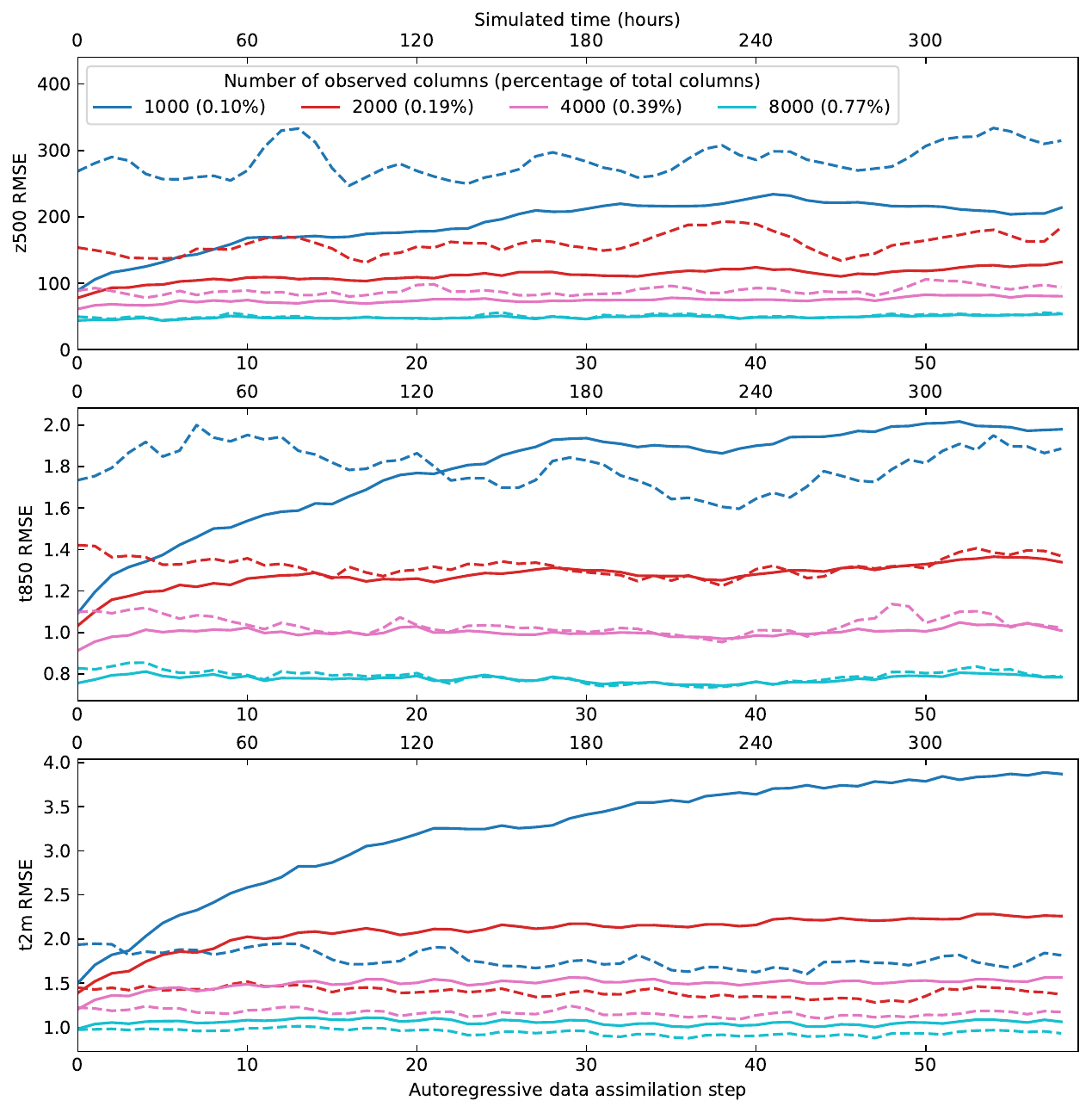}
    \caption{Root mean square errors (RMSEs) of geopotential at 500hPa, temperature at 850hPa and
temperature at 2m from autoregressively assimilated data (solid lines) and interpolated observations (dashed lines). The observations are made at the same locations in each assimilation step.}
    \label{fig:exp2_plot_fixed}
\end{figure}

\begin{figure}[h!]
    \centering
    \includegraphics[width=0.85\linewidth]{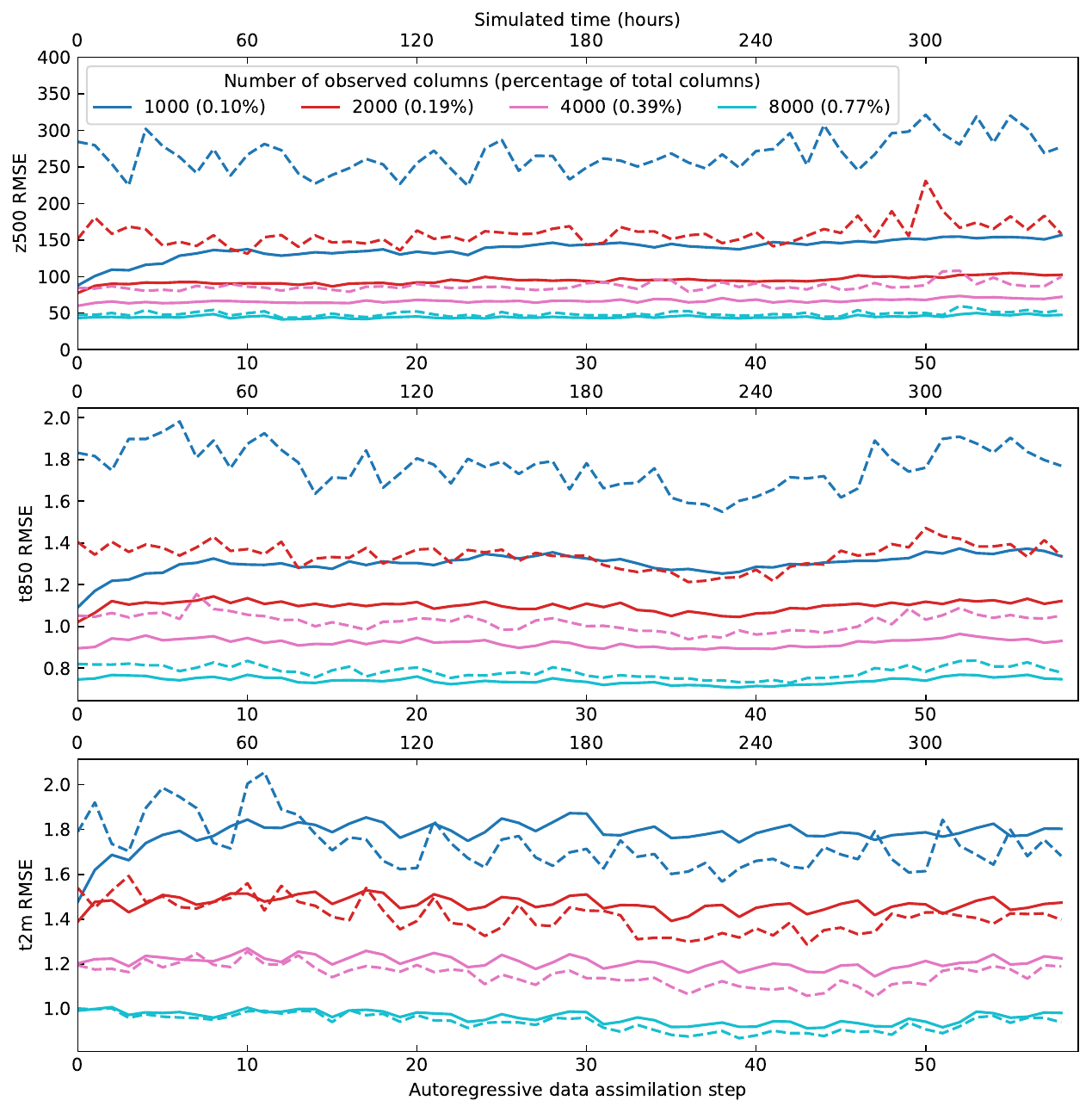}
    \caption{Root mean square errors (RMSEs) of geopotential at 500hPa, temperature at 850hPa and
temperature at 2m from autoregressively assimilated data (solid lines) and interpolated observations (dashed lines). The observations are made at different locations in each assimilation step.}
    \label{fig:exp2_plot}
\end{figure}

\paragraph{Autoregressive Data Assimilation} In an operational system, data assimilation is performed in an autoregressive way: assimilated data from previous time steps are used for producing forecast at the next time step, which is in turn used for assimilating data at that time step. Following this setting, we perform an autoregressive data assimilation cycle:
it starts from a 48-hour forecast at 0 hour, performs data assimilation at 0 hour, then performs a 6-hour forecast at 6 hour, and repeats the data assimilation -- 6-hour forecast cycle autoregressively. We train two separate diffusion models for this task, one dedicated to 48-hour forecast inputs and the other dedicated to 6-hour forecast inputs. In the data assimilation cycle, two strategies of sampling the simulated observation columns are used. One is to sample the columns at the same locations (\autoref{fig:exp2_plot_fixed}). The other resamples data at different locations in each iteration (\autoref{fig:exp2_plot}). The strategies help estimate the performance in real-world cases since the reality is the mix of the two strategies where some observations are measured at the same locations, such as weather stations and geostationary satellites, while locations of other observations change with time. We evaluate the autoregressive data assimilation result similar to the previous experiment. We compare the assimilation errors against errors of interpolated observations which gives a baseline for assimilated data.

We observe distinct patterns for fixed and non-fixed sampling strategies in the results. With fixed sampling  (\autoref{fig:exp2_plot_fixed}), the errors accumulates as the autoregressive data assimilation proceeds especially for fewer observations cases. The assimilation errors of z500 stabilize after several iterations and remain lower than the interpolation error. Meanwhile, assimilated t850 and t2m start from low errors and their errors grow gradually until exceeding the interpolation errors. In particular, our model struggles to assimilate t2m, where assimilated t2m deteriorate quickly and reach to higher error levels than interpolated observations.
In contrast, the non-fixed sampling strategy shows a slower error accumulation rate than fixed sampling (\autoref{fig:exp2_plot}). The assimilation errors of z500 and t850 remain lower than interpolation errors of observations throughout the autoregressive assimilation process. However, it is not the case for t2m where assimilation errors exceed interpolation errors after a few iterations.
This is likely due to the lack of four dimensional assimilation which can improve accuracy and stabilize the temporal roll out. We are working on four dimensional assimilation extension to our method and eventually target for assimilating real-world observations.

\paragraph{Forecast on Single-step Assimilated Data} Since assimilated data are often used as the input of weather forecasts, it is important to test the forecast errors using our assimilated data. In this experiment, we perform data assimilation at 0 hour, perform a 48-hour prediction using GraphCast, and then compare the resulting errors against the forecast errors with varying lead times using ERA5 as inputs. 

\begin{figure}[h!]
    \centering
    \includegraphics[width=\linewidth]{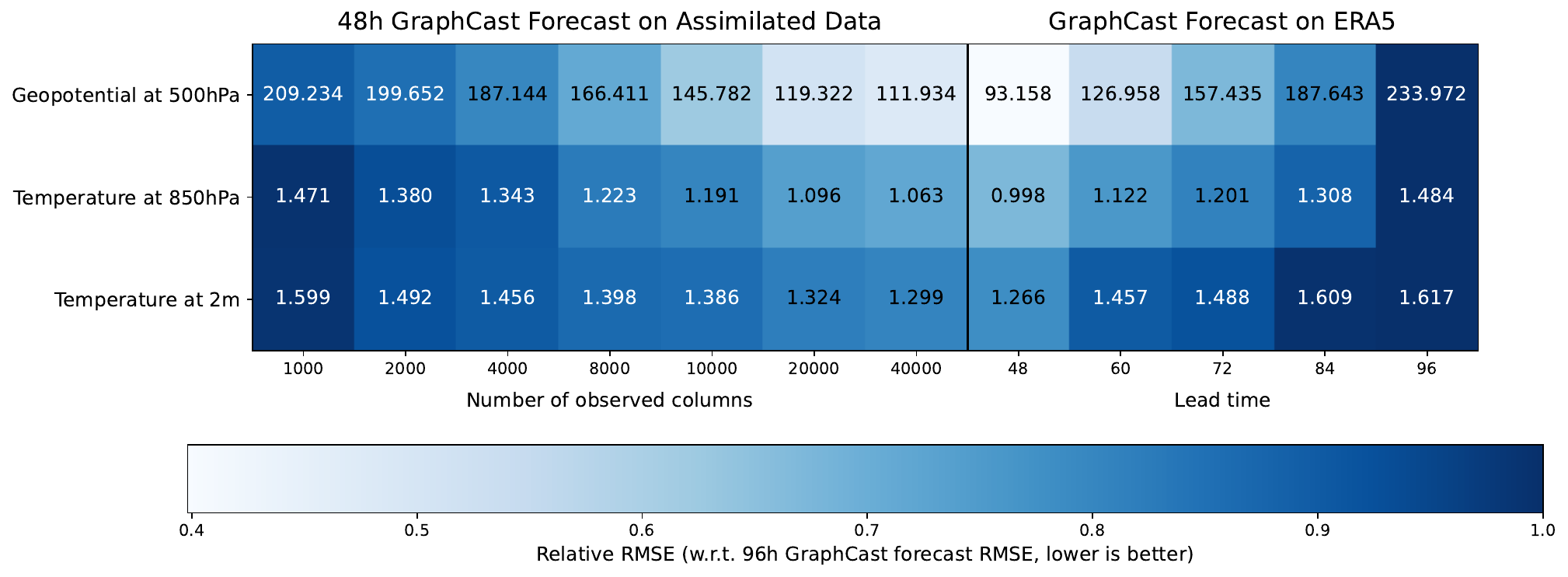}
    \caption{Root mean square errors (RMSEs, shown by the numbers in the cell) of geopotential at 500hPa, temperature at 850hPa, and temperature at 2m from the 48-hour forecast using assimilated data as inputs, and from forecasts with lead times from 48-hour to 96-hour using ERA5 as inputs. The errors are calculated against the ERA5 data. The cells are color-coded with the RMSEs relative to the 96-hour forecast errors.}
    \label{fig:exp3_score}
\end{figure}


As the scoreboard plot in \autoref{fig:exp3_score} shows, results from this experiment reveal that the 48-hour forecast errors using assimilated data inputs gradually converge to the 48-hour forecast errors using ERA5 inputs. More results can be found in \autoref{sec:scoreboard2}. The errors using assimilated data cannot be lower than using ERA5 inputs because the ERA5 data are used to simulate observations in the experiment. However, this setup allows us to compare the forecast errors against ones with longer lead times to determine the extent of lead time lost due to the use of assimilated data as forecast inputs. With 10,000 observed columns, the errors for z500, t850, and t2m are lower than the 72-hour forecast errors. This implies that the lead time lost is less than 24 hours for these variables. With 40,000 observed columns, the errors are lower than the 60-hour forecast errors. 




\section{Ablation Study}
We perform an ablation study on the Gaussian kernel standard deviation $\sigma_G$ to evaluate the effectiveness of the soft mask technique for incorporating sparse observations. A larger $\sigma_G$ corresponds to a larger effective range of the observation values. The soft mask degenerates into a hard mask when $\sigma_G=0$. We replicate the experiment settings from `single-step data assimilation' and `forecast on single-step data assimilation', varying $\sigma_G$ from $0.5$ to $3.0$. 

The ablation results for `forecast on single-step data assimilation' are shown in \autoref{tab:ablation}. Results for `single-step data assimilation' can be found in Appendix \autoref{tab:ablation2}. We observe an increase in accuracy when $\sigma_G$ grows from 0.5, demonstrating the advantage of the soft mask technique. The advantage likely comes from the growing region of the masked areas whereas the small masked areas in the hard mask can be smeared out by down-sampling layers in the diffusion model.
The errors saturate and increase again for $\sigma_G$ between 2.0 and 3.0 depending on the variable. 
In particular, t2m favors larger $\sigma_G$ than z500 and t850. Note that including too many interpolated values will also include their errors. In our main experiments, we chose $\sigma_G =2.5$.

\begin{table}[h!]
\centering
\caption{Root mean square errors (RMSEs) of geopotential at 500hPa, temperature at 850hPa, and temperature at 2m from the 48-hour forecast using assimilated data as inputs varying $\sigma_G$ and number of emulated observed columns.}
\label{tab:ablation}
\vspace{0.254cm}
\fontsize{9pt}{9pt}\selectfont
\begin{tabular}{rlllllll}
\toprule
& & \multicolumn{6}{c}{$\sigma_G$} \\ \cline{3-8}
&\textbf{\#Col} & 0.5 & 1.0 & 1.5 & 2.0 & 2.5 & 3.0 \\
\midrule
\parbox[t]{4mm}{\multirow{7}{*}{\rotatebox[origin=c]{90}{z500}}}\vline  &1000  & 224 & 216 & 209 & \textbf{200} & 209 & 212 \\
\vline  &2000  & 214 & 207 & 191 & \textbf{189} & 200 & 200 \\
\vline  &4000  & 211 & 194 & \textbf{177} & 180 & 187 & 181 \\
\vline  &8000  & 189 & 173 & 162 & 157 & 166 & \textbf{151} \\
\vline  &10000 & 180 & 164 & 148 & 147 & 146 & \textbf{142} \\
\vline  &20000 & 166 & 143 & 132 & 120 & \textbf{119} & 130 \\
\vline  &40000 & 139 & 120 & 115 & 111 & 112 & \textbf{105} \\
\midrule
\parbox[t]{4mm}{\multirow{7}{*}{\rotatebox[origin=c]{90}{t850}}}\vline &1000  & 1.53 & 1.49 & 1.46 & \textbf{1.42} & 1.47 & 1.46 \\
\vline &2000  & 1.47 & 1.44 & 1.40 & \textbf{1.38} & 1.38 & 1.43 \\
\vline &4000  & 1.47 & 1.39 & 1.32 & \textbf{1.31} & 1.34 & 1.32 \\
\vline &8000  & 1.37 & 1.27 & 1.22 & \textbf{1.21} & 1.22 & 1.21 \\
\vline &10000 & 1.33 & 1.26 & 1.18 & \textbf{1.16} & 1.19 & 1.18 \\
\vline &20000 & 1.26 & 1.14 & 1.11 & \textbf{1.09} & 1.10 & 1.12 \\
\vline &40000 & 1.14 & 1.07 & 1.07 & 1.06 & 1.06 & \textbf{1.04} \\
\midrule
\parbox[t]{4mm}{\multirow{7}{*}{\rotatebox[origin=c]{90}{t2m}}}\vline &1000  & 1.67 & 1.64 & 1.61 & 1.56 & 1.60 & \textbf{1.55} \\
\vline &2000  & 1.63 & 1.59 & 1.55 & 1.56 & \textbf{1.49} & 1.55 \\
\vline &4000  & 1.63 & 1.54 & 1.48 & 1.49 & \textbf{1.46} & 1.47 \\
\vline &8000  & 1.54 & 1.46 & 1.42 & 1.41 & \textbf{1.40} & 1.41 \\
\vline &10000 & 1.50 & 1.43 & 1.42 & \textbf{1.37} & 1.39 & 1.40 \\
\vline &20000 & 1.43 & 1.36 & 1.35 & 1.33 & \textbf{1.32} & 1.33 \\
\vline &40000 & 1.36 & 1.31 & 1.31 & 1.30 & 1.30 & \textbf{1.29} \\
\bottomrule
\end{tabular}
\end{table}
\section{Related Works}

Fengwu-4DVar~\citep{xiao2024fengwu4dvar} achieves 4-dimensional data assimilation with 1.4$^\circ$ resolution using the traditional 4D variational method. It utilizes the differentiable ML weather forecasting model Fengwu. Score-based data assimilation~\citep{rozet2023scoreDA} utilizes a diffusion model for data assimilation. It uses an additional loss function to enforce the observation values. It has been applied to assimilate 2D geospheric flows ~\citep{rozet2023scoreDAGeostrophic}.

MetNet3~\citep{andrychowicz2023metnet3} can make regional weather forecasts from gridded weather data and sparse observations. It combines data assimilation with prediction into one end-to-end process which requires training with sparse observation data. MetNet3 is susceptible to the generalization problem when training with observation data, thus it is hard to scale to a global high resolution scenario. GenCast~\citep{price2023gencast} is a diffusion model dedicated to probabilistic global weather forecast. It contains GraphCast-like encoder and decoder, and a transformer based processor. SEEDS~\citep{li2023seeds} is a diffusion model for generative emulation of weather forecast ensembles. It is capable of post-processing weather forecasts into ensembles which better represent the underlying probabilistic distribution of future weather states.

\section{Limitations and Future Works} 
Currently, our method can consume point measurements at a single time step as observations (\autoref{sec:prob_formulation}). This covers weather station measurements, upper-air balloon measurements, aircraft measurements, and satellite retrievals but does not include satellite imagery and radar soundings which can be represented as functions of the atmospheric states. We are planning to support broader types of observations and multiple time steps using loss-function-based conditioning at the inference period~\citep{chung2022dps, rozet2023scoreDA}. In addition, we are going to explore adding conditioning of the satellite imagery directly into the diffusion model similar to the conditioning of the predicted state (\autoref{sec:cond_pred}). This requires much more training effort but provides more flexibility as the observation function is learned implicitly. With more input observations, we expect to have more accurate and stable results when performing autoregressive data assimilation.

On the other hand, our method lacks quality control of the input observation data. It takes the input observations as ground truth values and enforces assimilated data to be consistent with the observations. Therefore, we have to resort to simulated observations from the ERA5 dataset. We are going to investigate possible quality control techniques that either pre-process the data separately or be embedded in the data assimilation process. By addressing the above points, we will achieve a fully operational data assimilation model.


\section{Conclusion}



In summary, we propose an easy-to-use data assimilation method, DiffDA, capable of assimilating high-resolution atmospheric variables up to a horizontal resolution of 0.25$^\circ$, setting a new record for ML-based data assimilation models. Based on the denoising diffusion model, we adapt the pre-trained GraphCast neural network as the backbone model motivated by its compatible input and output shapes. DiffDA's flexibility also allows for integrating other forecast models, ensuring easy updates and maintenance. 
A key feature of DiffDA is the conditioning on sparse observations occurs exclusively during inference which avoids ``the curse of dimensionality'' of adding sparse observations during training. An additional benefit of this conditioning approach is the automatic creation of a post-processing model, should observations not be supplied at inference time.

The experimental results validated the effectiveness of our method: the assimilated data converge to the observations as the number of observed data points increases. With observations occupying less than $0.96\%$ of total grid points and 48-hour forecast, the errors of assimilated data are on par with 24-hour forecast errors. When used as an input for forecast models, those assimilated data resulted in a maximum lead time loss of 24 hours compared to using ERA5 as inputs. This enables running data assimilation and simulation in an autoregressive cycle. It remains a challenge to constrain the errors across the autoregressive iterations. 

All the data assimilation experiments can run on a single high-end PC with a GPU within 15 - 30 minutes per data assimilation step, while a similar task using traditional methods typically requires large compute clusters. This indicates a significant reduction in computational costs. It also opens up the possibility of assimilating more observational data that is otherwise discarded by traditional methods, producing assimilated data with a higher accuracy.


\section*{Acknowledgements}
This study received funding from the MAELSTROM project funded by the European High-Performance Computing Joint Undertaking (JU; grant agreement No 955513).

\section*{Impact Statement}
This work aims for a more efficient and accurate data assimilation method, which will potentially help reduce the energy costs of weather forecast centers, make better weather forecasts, and provide better data for tackling climate change.

\bibliography{refs} 


\appendix
\section{Appendices}
\subsection{Code Availability}
The source code is available in this repository \href{https://github.com/spcl/DiffDA}{https://github.com/spcl/DiffDA}. The repository includes a modified GraphCast model which has the Apache License version 2.0.

\subsection{Scoreboard plots for single-step data assimilation}\label{sec:scoreboard1}

\begin{figure}[h!]
    \centering
    \includegraphics[width=\linewidth]{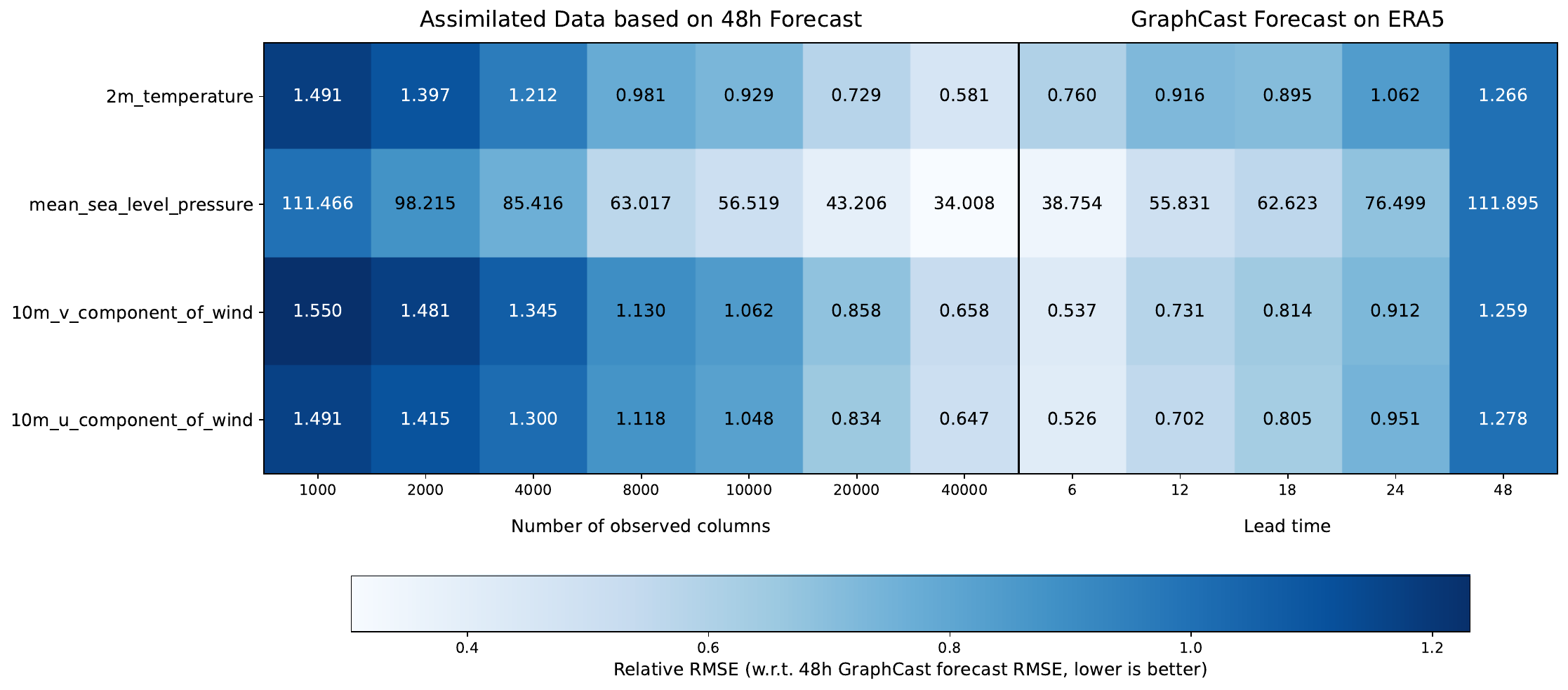}
    \caption{Root mean square errors (RMSEs, shown by the numbers in the cell) of surface variables from the single-step assimilated data, and from 6-hour to 48-hour GraphCast forecasts. The errors are calculated against the ERA5 data. The cells are color-coded with the RMSEs relative to the 48-hour forecast errors.}
\end{figure}

\begin{figure}[h!]
    \centering
    \includegraphics[width=\linewidth]{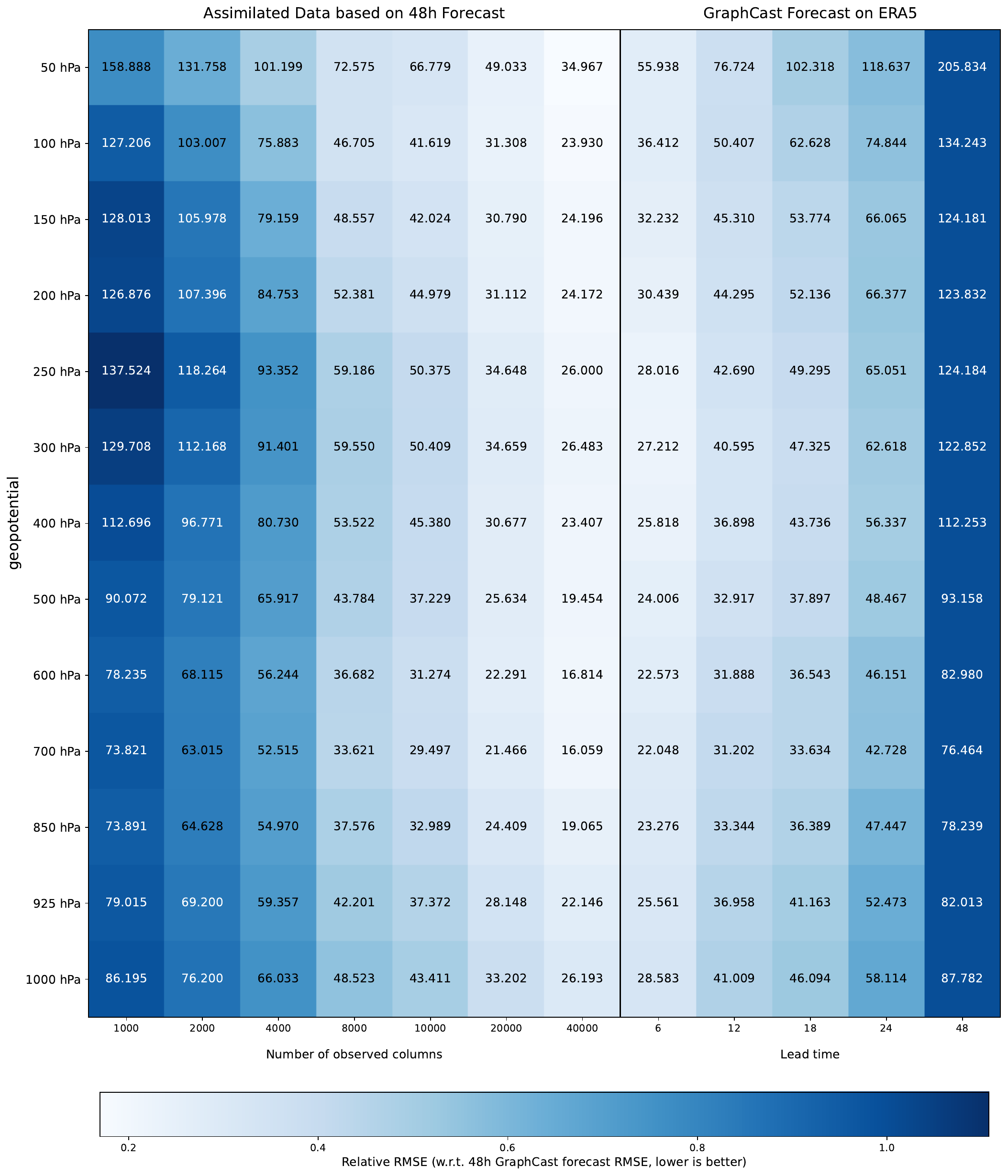}
    \caption{Root mean square errors (RMSEs, shown by the numbers in the cell) of geopotential from the single-step assimilated data, and from 6-hour to 48-hour GraphCast forecasts. The errors are calculated against the ERA5 data. The cells are color-coded with the RMSEs relative to the 48-hour forecast errors.}
\end{figure}

\begin{figure}[h!]
    \centering
    \includegraphics[width=\linewidth]{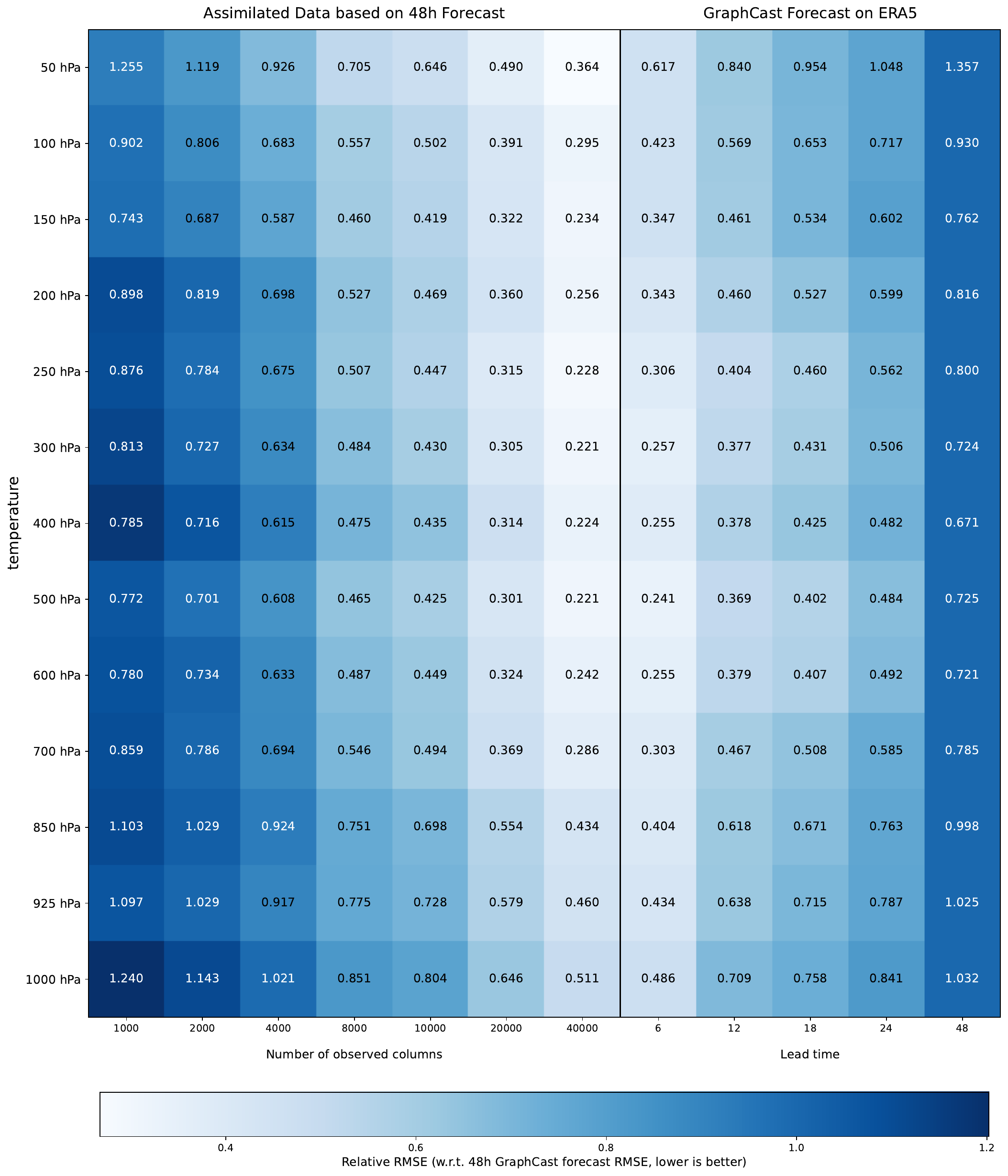}
    \caption{Root mean square errors (RMSEs, shown by the numbers in the cell) of temperature from the single-step assimilated data, and from 6-hour to 48-hour GraphCast forecasts. The errors are calculated against the ERA5 data. The cells are color-coded with the RMSEs relative to the 48-hour forecast errors.}
\end{figure}

\begin{figure}[h!]
    \centering
    \includegraphics[width=\linewidth]{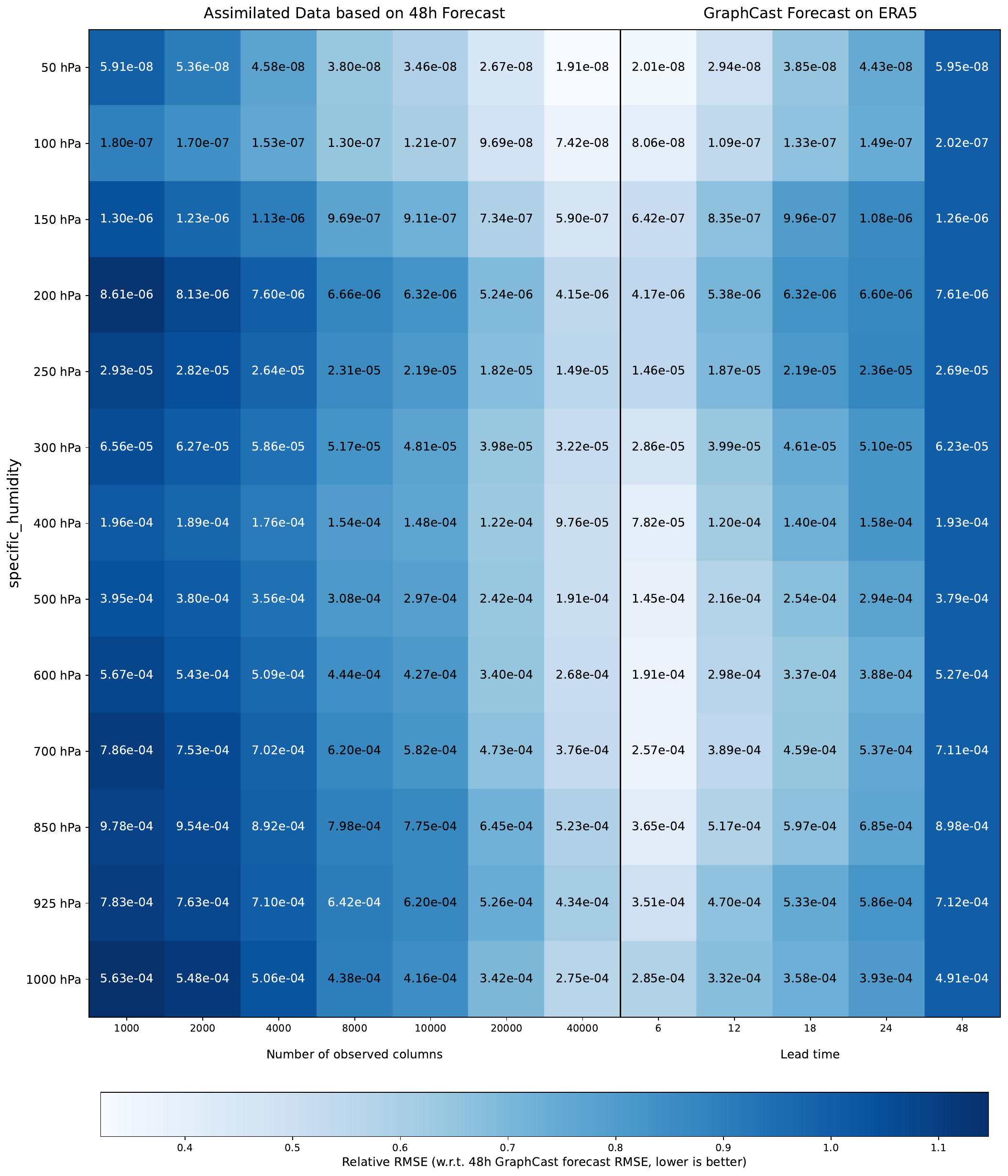}
    \caption{Root mean square errors (RMSEs, shown by the numbers in the cell) of specific humidity from the single-step assimilated data, and from 6-hour to 48-hour GraphCast forecasts. The errors are calculated against the ERA5 data. The cells are color-coded with the RMSEs relative to the 48-hour forecast errors.}
\end{figure}

\begin{figure}[h!]
    \centering
    \includegraphics[width=\linewidth]{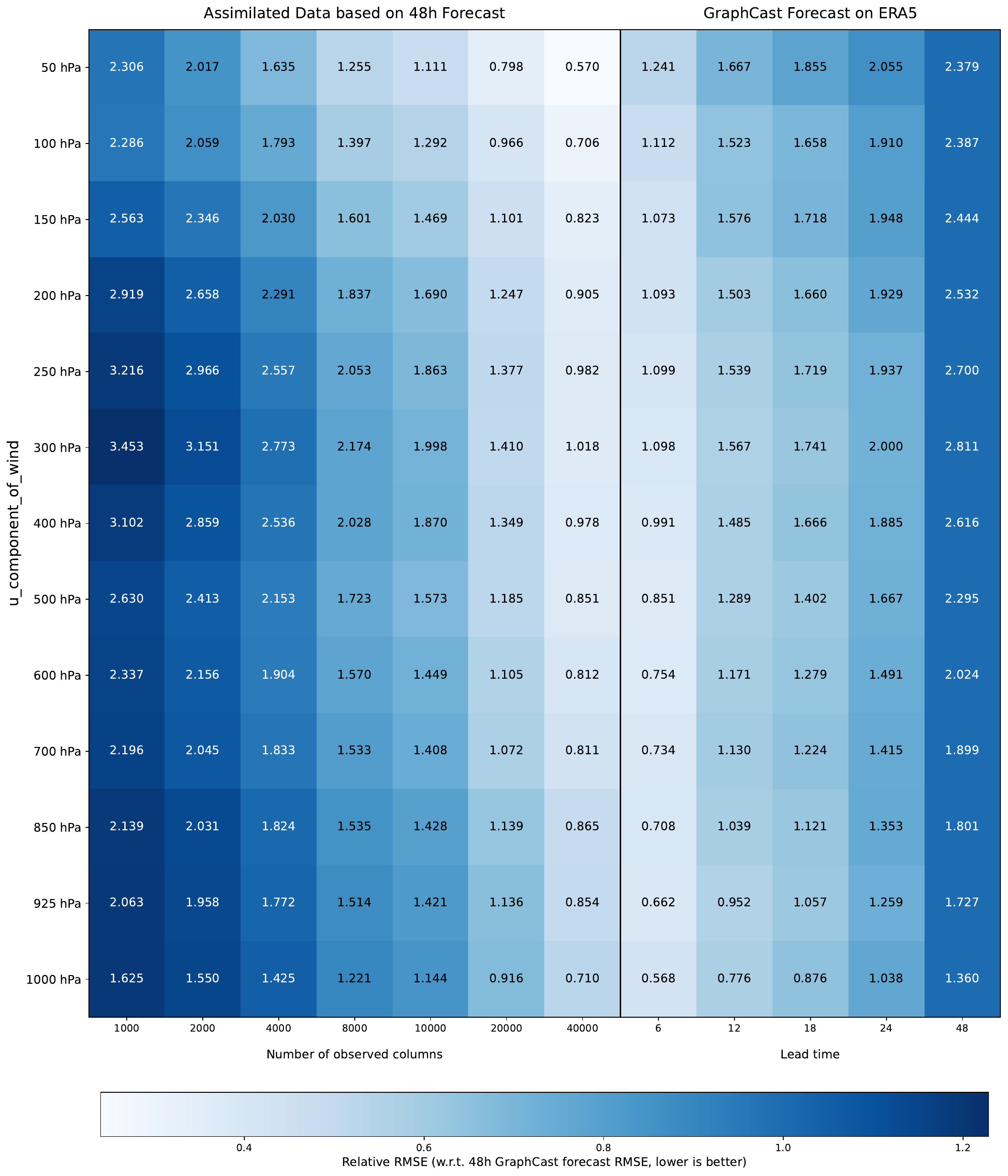}
    \caption{Root mean square errors (RMSEs, shown by the numbers in the cell) of horizontal wind speed (U) from the single-step assimilated data, and from 6-hour to 48-hour GraphCast forecasts. The errors are calculated against the ERA5 data. The cells are color-coded with the RMSEs relative to the 48-hour forecast errors.}
\end{figure}

\begin{figure}[h!]
    \centering
    \includegraphics[width=\linewidth]{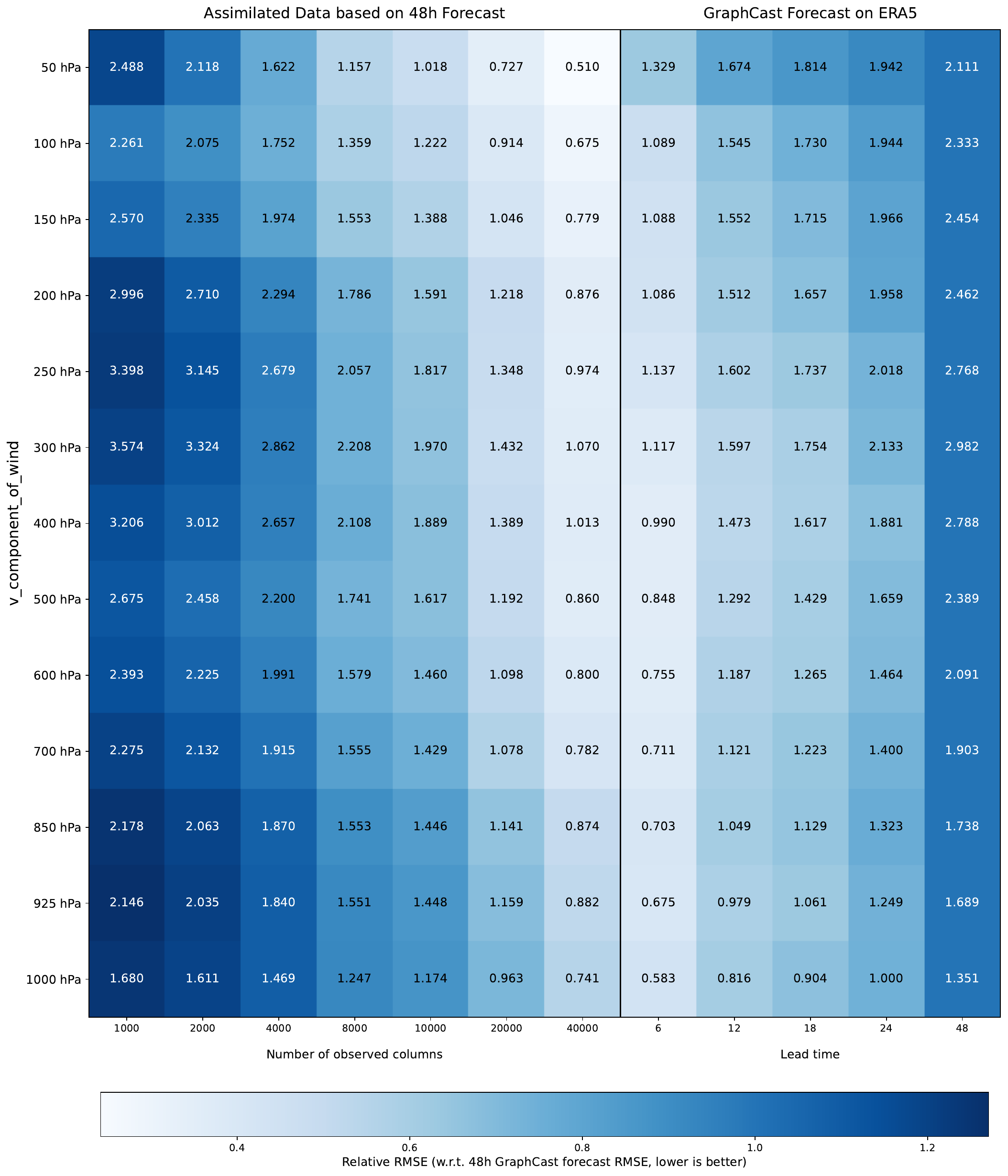}
    \caption{Root mean square errors (RMSEs, shown by the numbers in the cell) of horizontal wind speed (V) from the single-step assimilated data, and from 6-hour to 48-hour GraphCast forecasts. The errors are calculated against the ERA5 data. The cells are color-coded with the RMSEs relative to the 48-hour forecast errors.}
\end{figure}

\begin{figure}[h!]
    \centering
    \includegraphics[width=\linewidth]{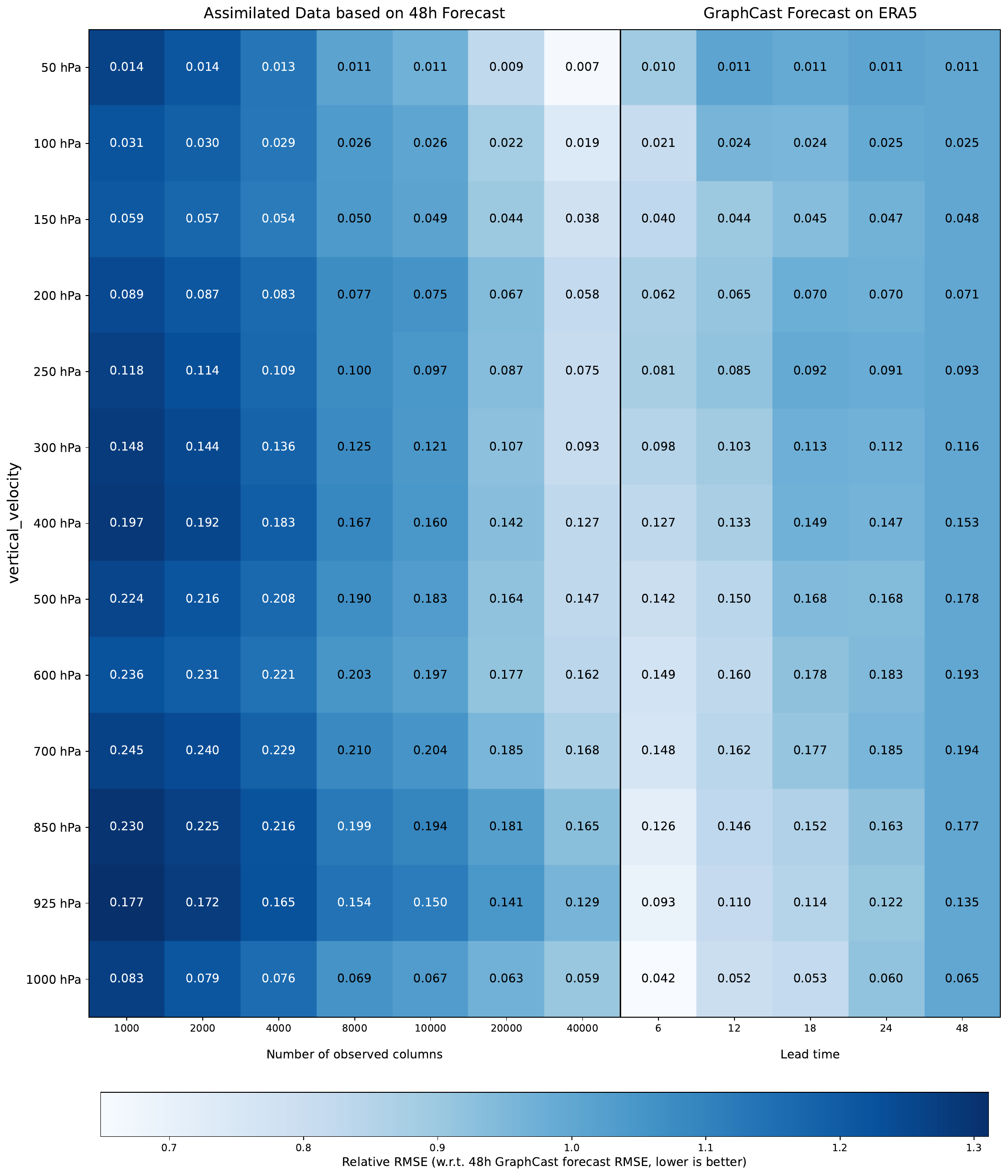}
    \caption{Root mean square errors (RMSEs, shown by the numbers in the cell) of vertical wind speed from the single-step assimilated data, and from 6-hour to 48-hour GraphCast forecasts. The errors are calculated against the ERA5 data. The cells are color-coded with the RMSEs relative to the 48-hour forecast errors.}
\end{figure}

\FloatBarrier
\subsection{Scoreboard plots for forecast on single-step assimilated data}\label{sec:scoreboard2}
\begin{figure}[h!]
    \centering
    \includegraphics[width=\linewidth]{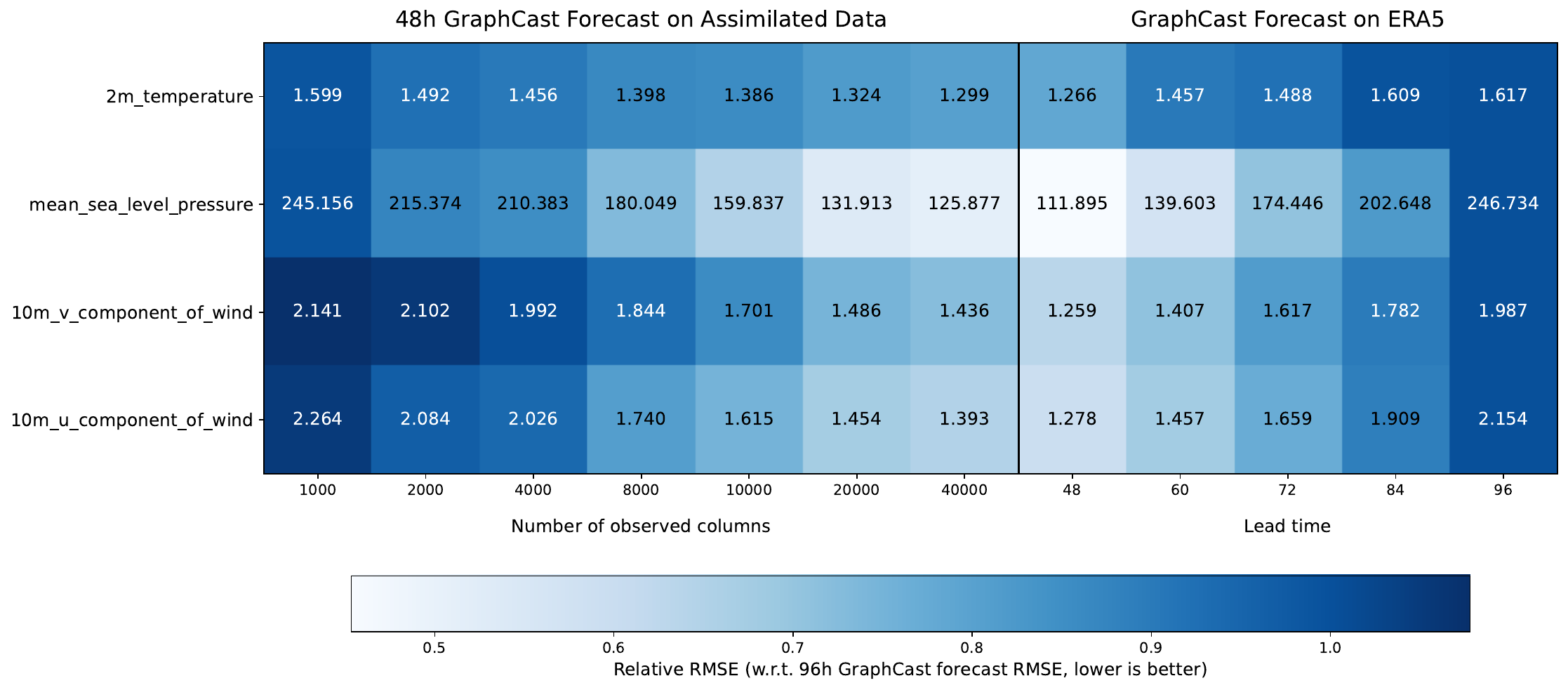}
    \caption{Root mean square errors (RMSEs, shown by the numbers in the cell) of surface variables from the 48-hour forecast using assimilated data as inputs, and from forecasts with lead times from 48-hour to 96-hour using ERA5 as inputs. The errors are calculated against the ERA5 data. The cells are color-coded with the RMSEs relative to the 96-hour forecast errors.}
\end{figure}

\begin{figure}[h!]
    \centering
    \includegraphics[width=\linewidth]{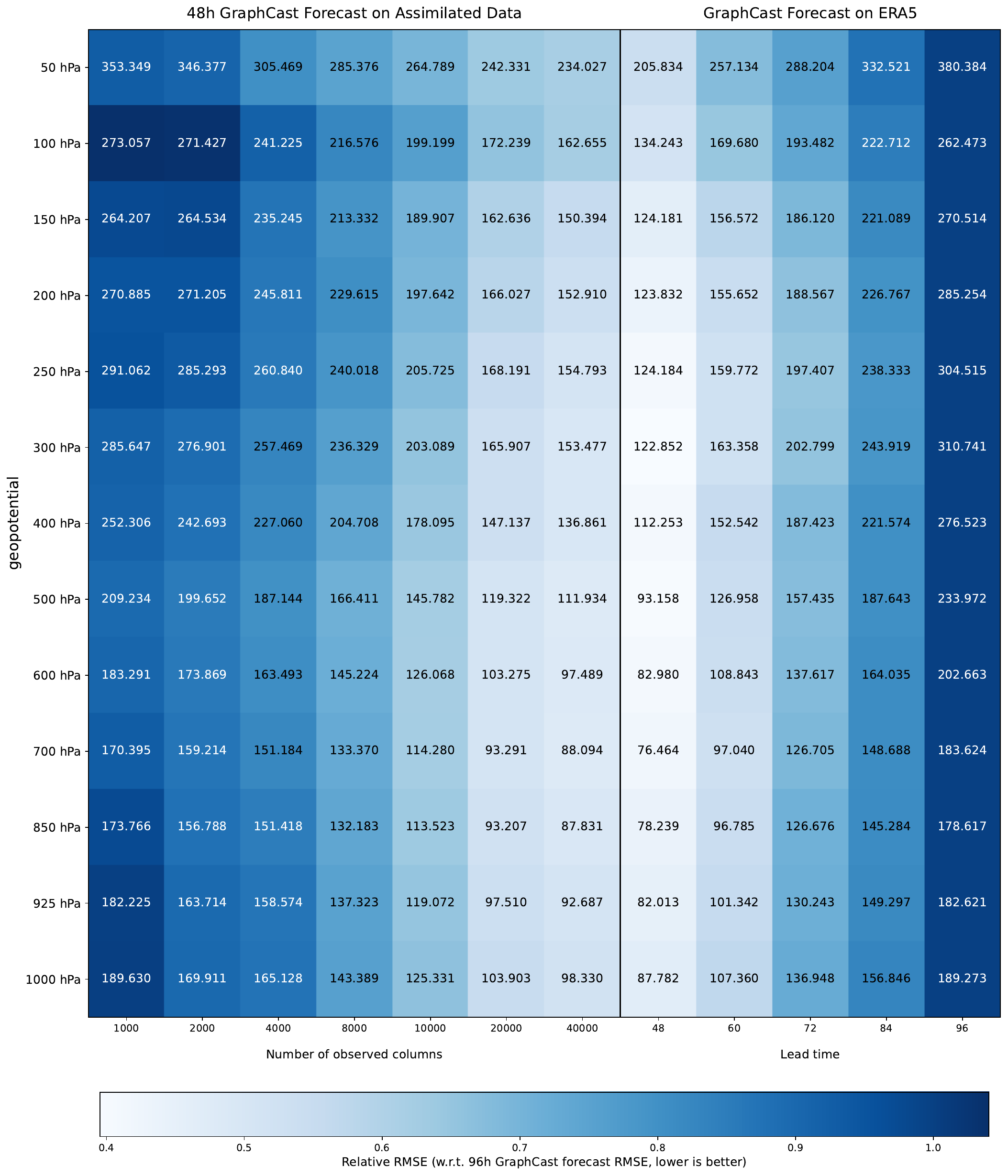}
    \caption{Root mean square errors (RMSEs, shown by the numbers in the cell) of geopotential from the 48-hour forecast using assimilated data as inputs, and from forecasts with lead times from 48-hour to 96-hour using ERA5 as inputs. The errors are calculated against the ERA5 data. The cells are color-coded with the RMSEs relative to the 96-hour forecast errors.}
\end{figure}

\begin{figure}[h!]
    \centering
    \includegraphics[width=\linewidth]{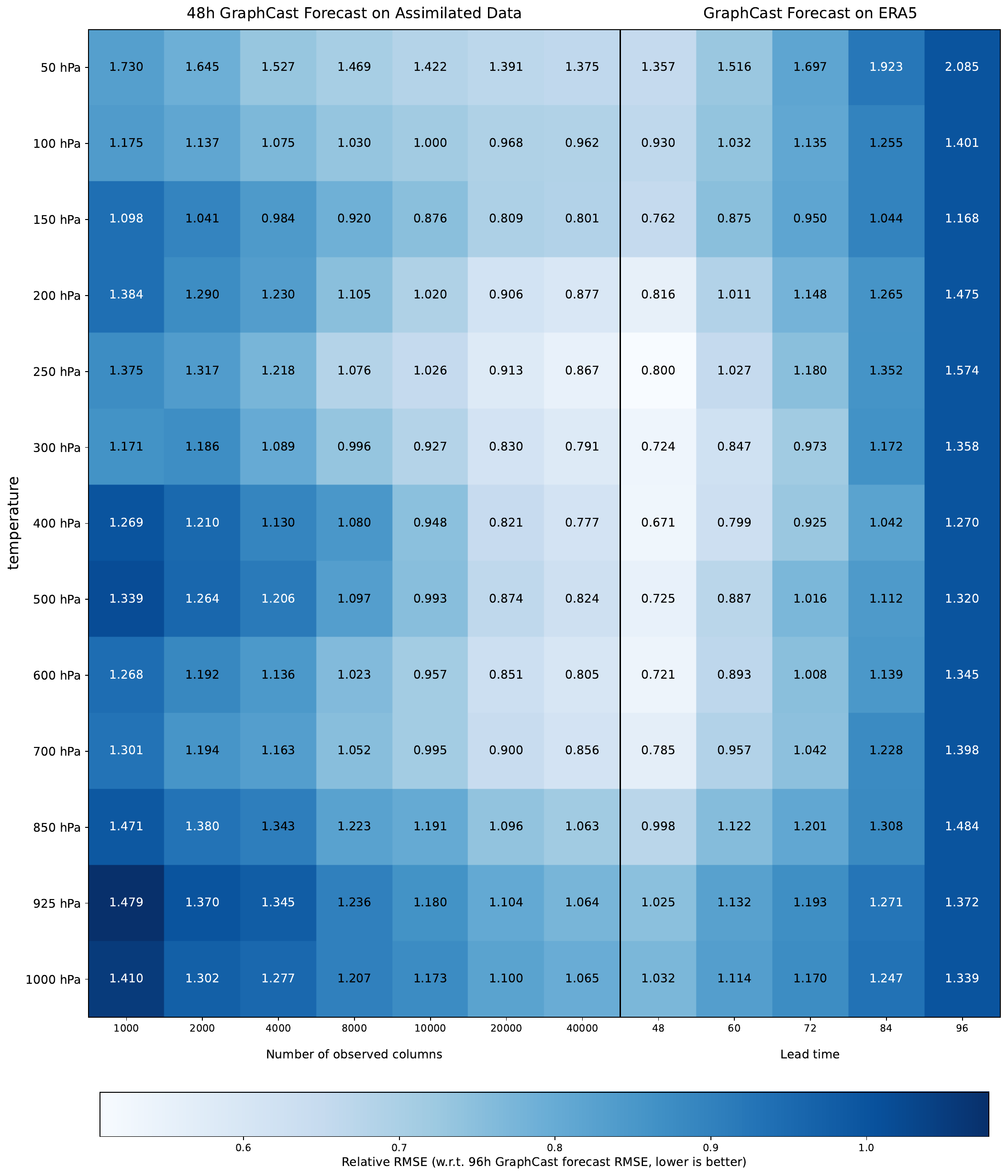}
    \caption{Root mean square errors (RMSEs, shown by the numbers in the cell) of temperature from the 48-hour forecast using assimilated data as inputs, and from forecasts with lead times from 48-hour to 96-hour using ERA5 as inputs. The errors are calculated against the ERA5 data. The cells are color-coded with the RMSEs relative to the 96-hour forecast errors.}
\end{figure}

\begin{figure}[h!]
    \centering
    \includegraphics[width=\linewidth]{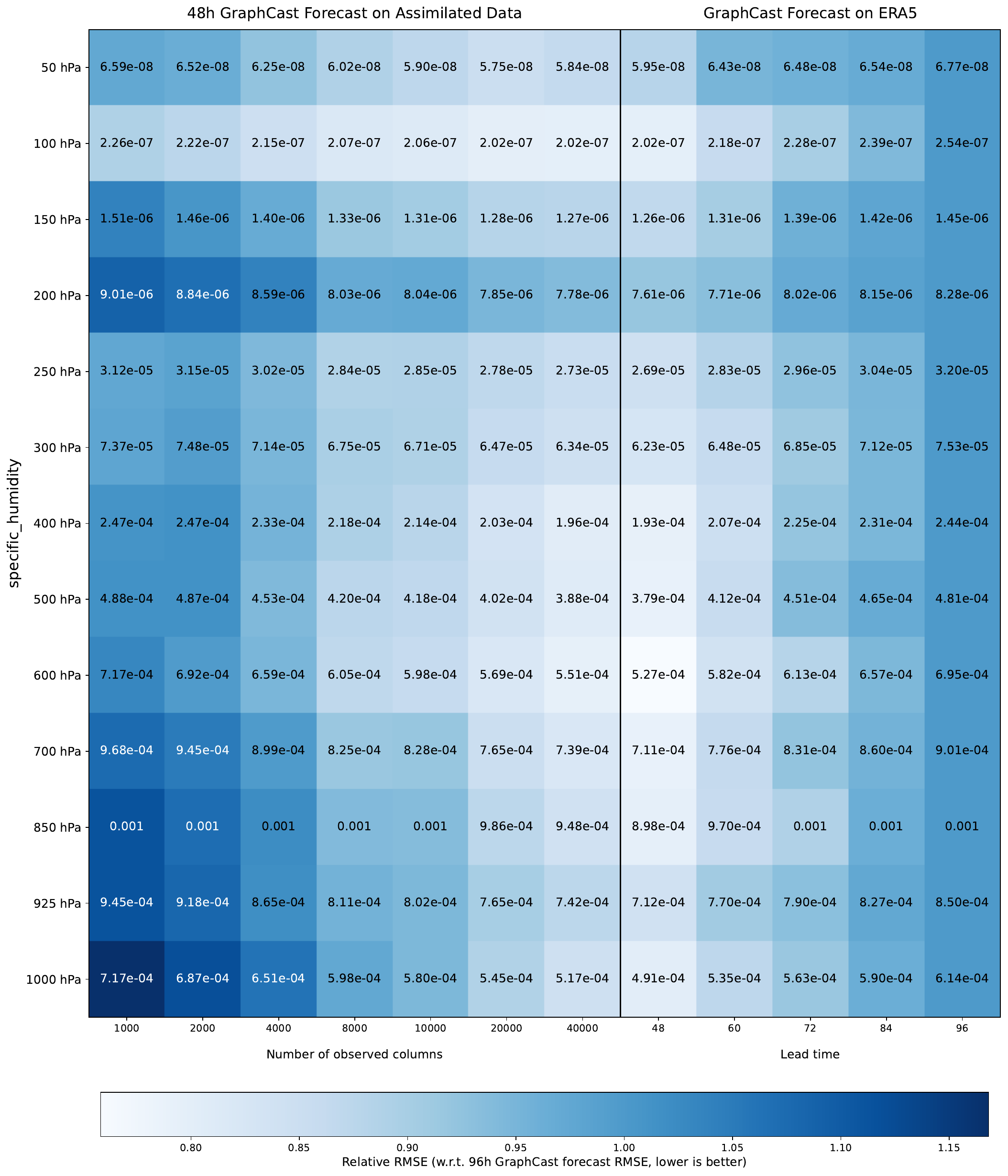}
    \caption{Root mean square errors (RMSEs, shown by the numbers in the cell) of specific humidity from the 48-hour forecast using assimilated data as inputs, and from forecasts with lead times from 48-hour to 96-hour using ERA5 as inputs. The errors are calculated against the ERA5 data. The cells are color-coded with the RMSEs relative to the 96-hour forecast errors.}
\end{figure}

\begin{figure}[h!]
    \centering
    \includegraphics[width=\linewidth]{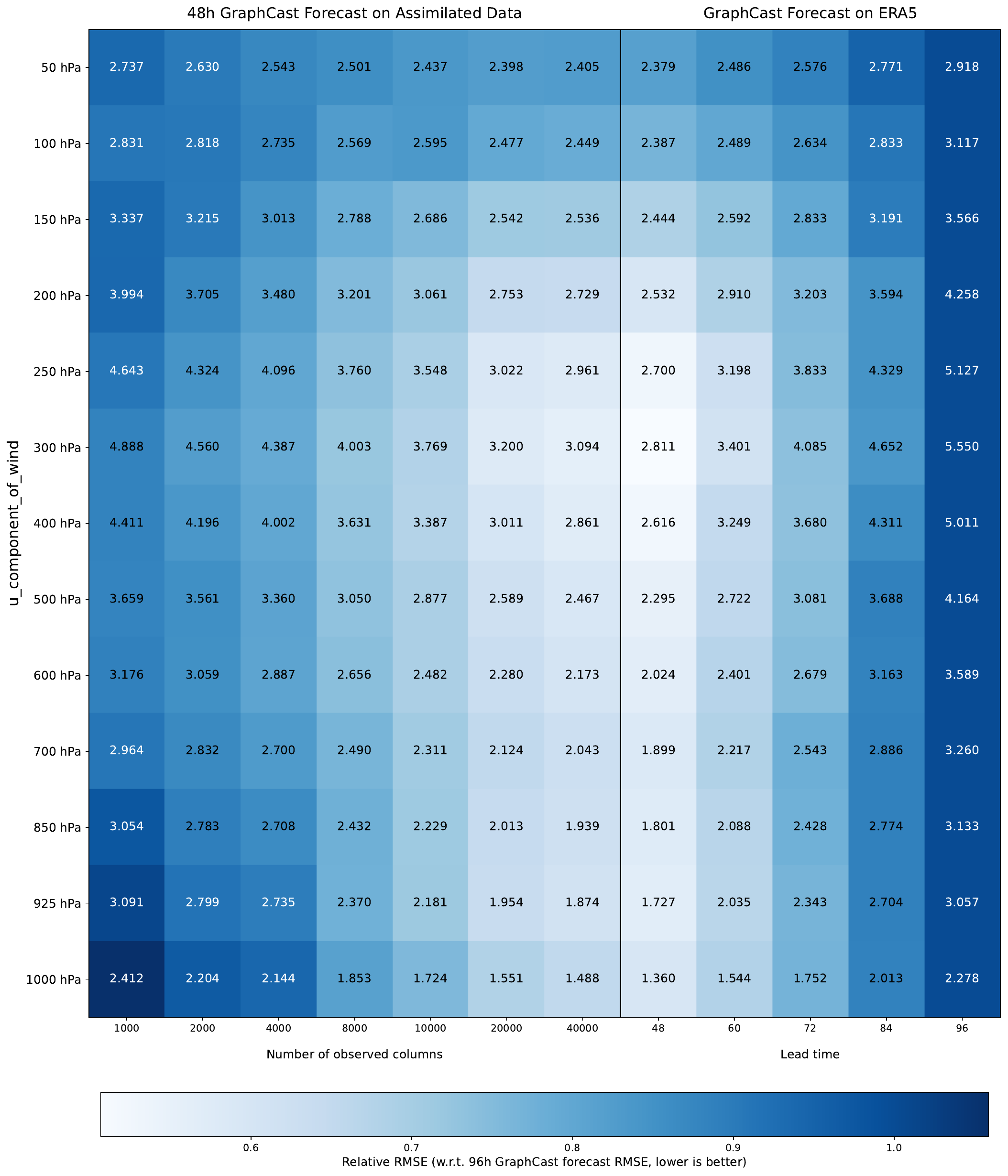}
    \caption{Root mean square errors (RMSEs, shown by the numbers in the cell) of horizontal wind speed (U) from the 48-hour forecast using assimilated data as inputs, and from forecasts with lead times from 48-hour to 96-hour using ERA5 as inputs. The errors are calculated against the ERA5 data. The cells are color-coded with the RMSEs relative to the 96-hour forecast errors.}
\end{figure}

\begin{figure}[h!]
    \centering
    \includegraphics[width=\linewidth]{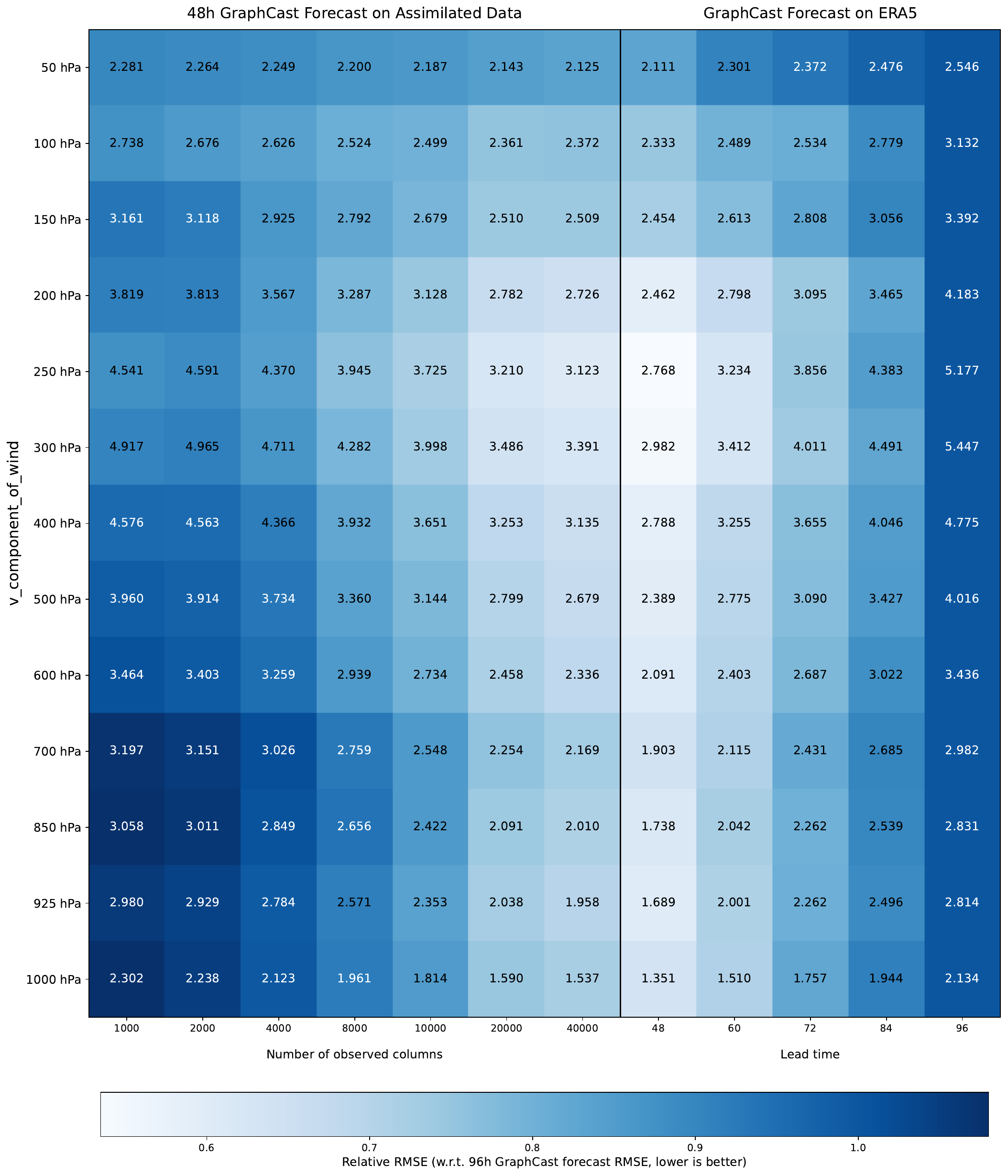}
    \caption{Root mean square errors (RMSEs, shown by the numbers in the cell) of horizontal wind speed (V) from the 48-hour forecast using assimilated data as inputs, and from forecasts with lead times from 48-hour to 96-hour using ERA5 as inputs. The errors are calculated against the ERA5 data. The cells are color-coded with the RMSEs relative to the 96-hour forecast errors.}
\end{figure}

\begin{figure}[h!]
    \centering
    \includegraphics[width=\linewidth]{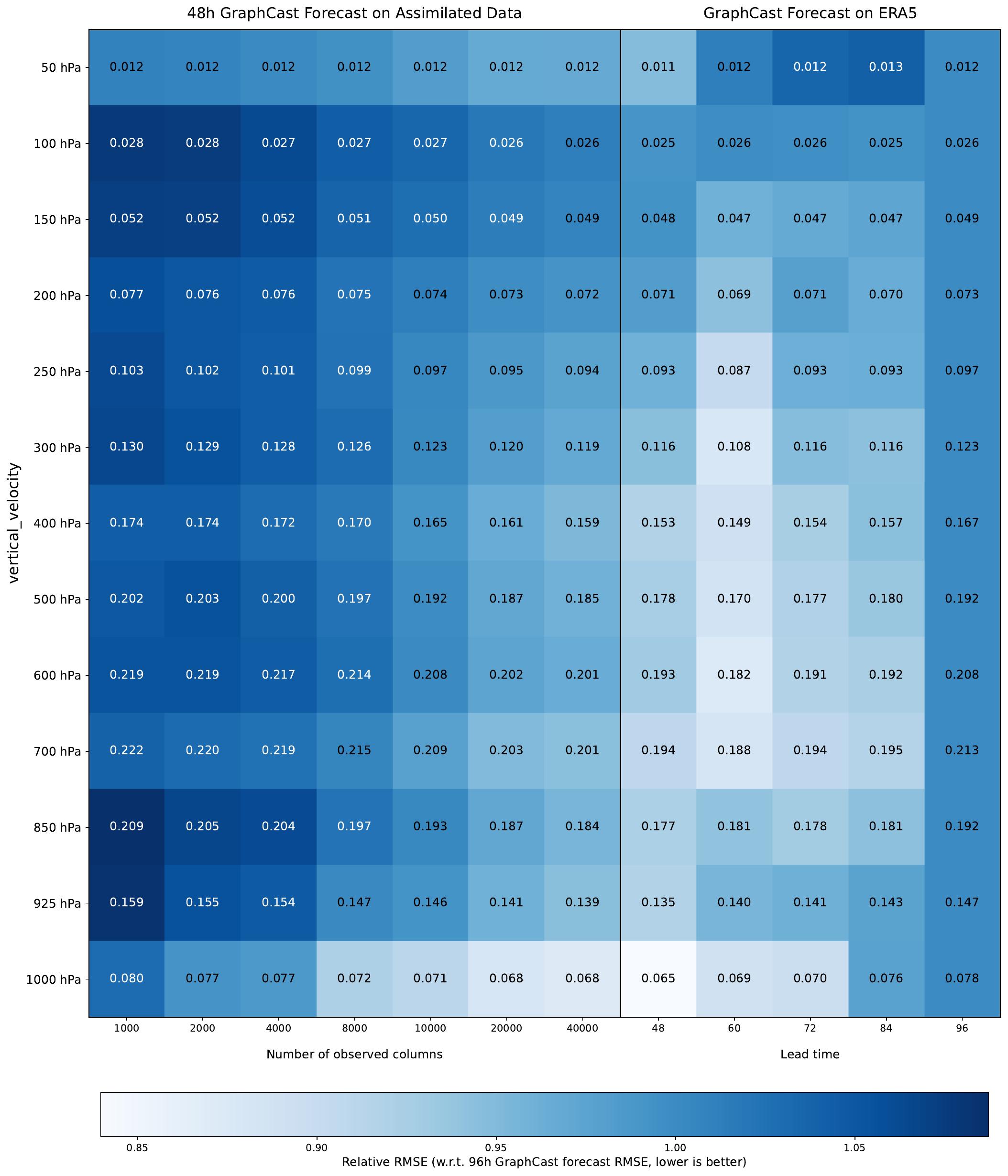}
    \caption{Root mean square errors (RMSEs, shown by the numbers in the cell) of vertical wind speed from the 48-hour forecast using assimilated data as inputs, and from forecasts with lead times from 48-hour to 96-hour using ERA5 as inputs. The errors are calculated against the ERA5 data. The cells are color-coded with the RMSEs relative to the 96-hour forecast errors.}
\end{figure}

\FloatBarrier
\subsection{Table for ablation study results}
\begin{table}[h!]
\centering
\caption{Root mean square errors (RMSEs) of geopotential at 500hPa, temperature at 850hPa, and temperature at 2m from the single-step assimilated data varying $\sigma_G$ and number of emulated observed columns.}
\label{tab:ablation2}
\vspace{1em}
\fontsize{9pt}{9pt}\selectfont
\begin{tabular}{rlllllll}
\toprule
& & \multicolumn{5}{c}{$\sigma_G$} \\ \cline{3-8}
&\textbf{\#Columns} & 0.5 & 1.0 & 1.5 & 2.0 & 2.5 & 3.0 \\
\midrule
\parbox[t]{4mm}{\multirow{7}{*}{\rotatebox[origin=c]{90}{z500}}}\vline  &1000  & 94.6 & 89.7 & 86.7 & \textbf{86.3} & 90.1 & 92.8 \\
\vline  &2000  & 90.7 & 83.0 & 78.0 & \textbf{76.3} & 79.1 & 81.0 \\
\vline  &4000  & 84.5 & 70.7 & 63.0 & 63.5 & 65.9 & \textbf{62.6} \\
\vline  &8000  & 72.8 & 54.9 & 47.0 & \textbf{43.0} & 43.8 & 43.0 \\
\vline  &10000 & 68.5 & 50.6 & 41.7 & 38.3 & 37.2 & \textbf{36.9} \\
\vline  &20000 & 52.2 & 35.0 & 28.8 & 27.1 & \textbf{25.6} & 26.7 \\
\vline  &40000 & 36.9 & 23.9 & 20.7 & 19.6 & 19.5 & \textbf{18.7} \\
\midrule
\parbox[t]{4mm}{\multirow{7}{*}{\rotatebox[origin=c]{90}{t850}}}\vline &1000  & 1.14 & 1.12 & 1.11 & \textbf{1.09} & 1.10 & 1.09 \\
\vline &2000  & 1.13 & 1.09 & 1.06 & 1.04 & 1.03 & \textbf{1.02} \\
\vline &4000  & 1.10 & 1.04 & 0.98 & 0.94 & 0.92 & \textbf{0.90} \\
\vline &8000  & 1.03 & 0.93 & 0.84 & 0.78 & 0.75 & \textbf{0.73} \\
\vline &10000 & 1.01 & 0.88 & 0.79 & 0.73 & 0.70 & \textbf{0.69} \\
\vline &20000 & 0.89 & 0.72 & 0.61 & 0.56 & \textbf{0.55} & 0.56 \\
\vline &40000 & 0.73 & 0.53 & 0.46 & \textbf{0.43} & 0.43 & 0.44 \\
\midrule
\parbox[t]{4mm}{\multirow{7}{*}{\rotatebox[origin=c]{90}{t2m}}}\vline &1000  & 1.58 & 1.57 & 1.53 & 1.51 & 1.49 & \textbf{1.47} \\
\vline &2000  & 1.55 & 1.51 & 1.48 & 1.42 & 1.40 & \textbf{1.35} \\
\vline &4000  & 1.52 & 1.43 & 1.35 & 1.28 & 1.21 & \textbf{1.19} \\
\vline &8000  & 1.44 & 1.31 & 1.18 & 1.07 & 0.98 & \textbf{0.95} \\
\vline &10000 & 1.40 & 1.26 & 1.11 & 0.98 & 0.93 & \textbf{0.89} \\
\vline &20000 & 1.27 & 1.04 & 0.85 & 0.76 & \textbf{0.73} & 0.74 \\
\vline &40000 & 1.08 & 0.77 & 0.62 & 0.60 & \textbf{0.58} & 0.59 \\
\bottomrule
\end{tabular}
\end{table}

\end{document}